\documentclass[conference]{IEEEtran}
\def\BibTeX{{\rm B\kern-.05em{\sc i\kern-.025em b}\kern-.08em
    T\kern-.1667em\lower.7ex\hbox{E}\kern-.125emX}}

\usepackage{mathtools} 
\usepackage{standalone}
\usepackage{tikz,url}
\usepackage{tikz-cd}
\usetikzlibrary{arrows,automata,decorations,decorations.pathmorphing,fit}
\usepackage{amsmath,amssymb, amsfonts, amsthm}
\usepackage{mathrsfs}
\usepackage{mathpartir}
\usepackage{tikzit}

\tikzstyle{white vertex}=[fill=white, draw=black, shape=circle, thick, inner sep=1pt, minimum size=6pt]
\tikzstyle{green vertex}=[fill=green, draw=black, shape=circle, thick, inner sep=1pt, minimum size=6pt]
\tikzstyle{cyan vertex}=[fill=cyan, draw=black, shape=circle, thick, inner sep=1pt, minimum size=6pt]
\tikzstyle{magenta vertex}=[fill={magenta!60}, draw=black, shape=circle, thick, inner sep=1pt, minimum size=6pt]
\tikzstyle{empty node}=[fill=white, shape=circle, inner sep=1pt]
\tikzstyle{small font vertex}=[fill=none, draw=none, shape=circle, font={{\scriptsize}}]
\tikzstyle{new style 0}=[fill=none, draw=none, shape=circle]

\tikzstyle{filled path}=[-, fill={blue!60}, thick, fill opacity=0.15]
\tikzstyle{thick edge}=[-, thick]
\tikzstyle{dashed path}=[-, dashed]
\tikzstyle{arrow path}=[->]

\usepackage{macros}
\usepackage{enumerate}
\usepackage[hidelinks]{hyperref}
\usepackage[capitalize]{cleveref}
\usepackage{wrapfig}

\def\ECn{\mathbf{EC_n}}

\newtheorem{proposition}{Proposition}
\newtheorem{lemma}{Lemma}
\newtheorem{theorem}{Theorem}
\newtheorem{corollary}{Corollary}

\theoremstyle{definition}
\newtheorem{definition}{Definition}
\newtheorem{remark}{Remark}
\newtheorem{example}{Example}

\title{Semi-simplicial Set Models for Distributed Knowledge}

\author{
\IEEEauthorblockN{\'Eric Goubault\IEEEauthorrefmark{1}, Roman Kniazev\IEEEauthorrefmark{1}\IEEEauthorrefmark{2}, J\'er\'emy Ledent\IEEEauthorrefmark{3}, Sergio Rajsbaum\IEEEauthorrefmark{4}}\\
\IEEEauthorblockA{\IEEEauthorrefmark{1}
	\textit{LIX, CNRS, \'Ecole Polytechnique, IP-Paris}, Palaiseau Cedex, France
}
\IEEEauthorblockA{\IEEEauthorrefmark{2}
	\textit{Université Paris-Saclay, ENS Paris-Saclay, CNRS, LMF}, Gif-sur-Yvette, France}
\IEEEauthorblockA{\IEEEauthorrefmark{3}
	\textit{MSP Group, University of Strathclyde}, Glasgow, Scotland
}
\IEEEauthorblockA{\IEEEauthorrefmark{4}
	\textit{Universidad Nacional Autónoma de México (UNAM)}, Mexico D.F., Mexico and 
	IRIF, Paris, France 
}
}

\IEEEoverridecommandlockouts
\IEEEpubid{\makebox[\columnwidth]{979-8-3503-3587-3/23/\$31.00~
\copyright2023 IEEE \hfill} \hspace{\columnsep}\makebox[\columnwidth]{ }}

\makeatletter
 \renewcommand\@oddhead{\thepage}
 \makeatother

\begin{document}
\maketitle

\begin{abstract}
In recent years, a new class of models for multi-agent epistemic logic has emerged, based on simplicial complexes. Since then, many variants of these simplicial models have been investigated, giving rise to different logics and axiomatizations.
In this paper, we present a further generalization, which encompasses all previously studied variants of simplicial models.
Geometrically, this is achieved by generalizing beyond simplicial complexes, and considering instead semi-simplicial sets.
By doing so, we define a new semantics for epistemic logic with distributed knowledge, where a group of agents may distinguish two worlds, even though each individual agent in the group is unable to distinguish them.
As it turns out, these models are the geometric counterpart of a generalization of Kripke models, called ``pseudo-models’’.
We show how to recover the previously defined variants of simplicial models as sub-classes of our models; and give a sound and complete axiomatization for each of them\let\thefootnote\relax
\footnote{Eric Goubault was partially funded by AID/CIEDS project FARO. Part of the work of Sergio Rajbaum was performed while he was an invited professor at IRIF, Université Paris Cité, and at LIX, Ecole Polytechnique.}.
\end{abstract}



\section{Introduction}
The usual semantics for  multi-agent epistemic logic is 
 based on the classic Kripke  \emph{possible worlds} relational structure~\cite{fagin}.
However, the intimate relationship between distributed computing and algebraic topology~\cite{herlihyetal:2013} showed the importance of moving from the focus on global states represented by worlds, to local states, representing {perspectives} about possible worlds. Namely, moving from graph structures to simplicial complex structures.
A  formal semantics of multi-agent epistemic formulas in terms of simplicial models was presented~\cite{gandalf-journal}, and shown to be equivalent to the usual Kripke model semantics for $\Sfive$. 
Further work explored bisimilarity of simplicial models~\cite{Ditmarsch2020KnowledgeAS} and connections with covering spaces~\cite{DitmarschGLLR21}.  Remarkably, it has been shown by Yagi and Nishimura~\cite{yagiNishimura2020TR,Nishimura22mucalculus} that the implicit topological information in Kripke models, exposed by the simplicial complex point of view, can be leveraged to produce a logical obstruction to the solvability of certain distributed computing problems.
The notion of \emph{distributed knowledge}~\cite{HalpernM90} plays a crucial role there: in a sense, it is a higher-dimensional notion of knowledge.

These first results assumed a finite, fixed set of agents, whose local states appear in every world.
As a consequence, since a world with $n+1$ agents is represented by a $n$-dimensional simplex, every facet of the simplicial model is of the same dimension.
Such models are called \emph{pure} simplicial models.
However, in distributed systems, processes may fail, and it may happen that only a subset of the agents remain in some worlds.
To model such situations, the categorical equivalence of~\cite{gandalf-journal} was extended by various authors to include simplicial models that may not be pure~\cite{GoubaultLR21kb4,Ditmarsch21Wanted,Ditmarsch22Complete}.
These situations have been thoroughly studied since early on in distributed computability, e.g.\ the  seminal work of Dwork and Moses~\cite{DworkM90crash}, where a complete characterization of the number of rounds required to reach simultaneous consensus was given in terms of common knowledge.
By moving away from pure simplicial models, we also move away from the standard $\Sfive$ epistemic logic.
Indeed, there are a number of design choices to be made, both to define the models and their semantics. Should all simplices represent worlds, or only the facets? Can we have a world with no agents? How do we deal with formulas involving the knowledge of agents that are not present in the current world?
All these choices can greatly influence the resulting logic: while~\cite{GoubaultLR21kb4} drops Axiom~\textbf{T} and works with the logic $\KBfour$ (augmented with some extra axioms), other authors~\cite{Ditmarsch22Complete} take a completely different route and define a three-valued logic (where formulas may be undefined), with an axiom system called $\textbf{S5}^{\bowtie}$.

In this paper, we aim to bring order to chaos by defining a class of simplicial models that encompasses all previous variants.
Thus, the main objective of our paper is to \textit{unify} previous work on simplicial complex models.
For that purpose, we introduce a new class of models called \emph{epistemic covering models}.
As we will see in \cref{sec:subclasses-literature}, the models studied in \cite{gandalf-journal, GoubaultLR21kb4, Ditmarsch21Wanted, baltagCorrelatedKnowledgeEpistemicLogic2010}
can all be viewed as sub-classes of epistemic covering models.
This allows us to recover important theoretical results about each of those sub-classes, by proving them once and for all in our very general setting.
Namely, our paper contains two central results:
(i) \cref{prop:main2} shows an equivalence of categories between epistemic covering models and their Kripke-style counterpart, called \emph{generalized epistemic models}.
Crucially, we also show (\cref{lem:properties-agree}) that this equivalence can safely be restricted to sub-classes of epistemic covering models; allowing us to recover similar equivalence results from \cite{gandalf-journal, GoubaultLR21kb4, Ditmarsch21Wanted}.
(ii) The other essential result of our paper is a sound and complete axiomatization of epistemic covering models (\cref{cor:axiomatization-coverings}).
Once again, we are also interested in axiomatizing the various sub-classes of epistemic covering models, which we do in \cref{sec:subclass-axioms}.
We show that there is a close correspondence between structural properties that define a sub-class, and axioms of the corresponding logic.
These results are summed up in \cref{fig:axiom-table} below.
The first two columns indicate the correspondence between properties of epistemic coverings, and axioms of the logic (they will be defined in \cref{sec:subclasses,sec:axiomatization}, respectively).
The check marks in the other columns indicate how to recover previously-studied variants of simplicial models.
One can check that the axiomatization provided here matches the one given in each of those papers (except for \cite{Ditmarsch21Wanted}, which studies a different three-valued semantics).

 \begin{figure}[!h]
 \begin{tabular}{llcccc} \hline
 Covering properties & Axioms & \cite{gandalf-journal} & \cite{GoubaultLR21kb4} & \cite{Ditmarsch21Wanted} & \cite{baltagCorrelatedKnowledgeEpistemicLogic2010} \\ \hline
 Proper & ($\mathbf{P}$) & \checkmark & \checkmark & \checkmark & \checkmark \\
 Pure & ($\mathbf{T}$) & \checkmark &  & & \checkmark \\
 Minimal & ($\mathbf{Min}$) & \checkmark & \checkmark & & \checkmark \\
 Maximal & ($\mathbf{Max}$) &  &  & \checkmark & \\
 No empty world & ($\mathbf{NE}$) & \checkmark & \checkmark & \checkmark & \checkmark \\
 Simplicial complex & no axiom & \checkmark & \checkmark & \checkmark \\ \hline
 \end{tabular}
 \caption{\label{fig:axiom-table} Sub-classes of epistemic covering models that have been studied in previous papers, and their associated axioms.}
 \end{figure}

One distinguishing feature of our models, which has not been considered previously in the literature, is that we generalize beyond simplicial complexes and consider instead semi-simplicial sets.
Geometrically, this means that simplices (a.k.a.\ worlds) can be connected in more complex ways; for instance, two triangles might share two vertices but not the edge between those vertices (see \cref{fig:sset-ex}, which is not a simplicial complex).
This new feature enables us to model situations where a group of agents may distinguish two worlds, even though each individual agent in the group is unable to distinguish them.

A toy example of such a situation, in the realm of distributed computing
is the following. 
Suppose we have three sensors, $s_1$, $s_2$ and $s_3$, in a sensor network, with overlapping visibility regions of the form of a unit disk, that can only count the number of targets within their visibility region. 
%
Suppose now that each of these three sensors detects exactly one target, indicated by a ``$\times$'' below. There are 5 possible configurations $w_1$, $w_2$, $\ldots$, $w_5$ (from left to right below), with the total number of targets, ranging from 1 to~3:
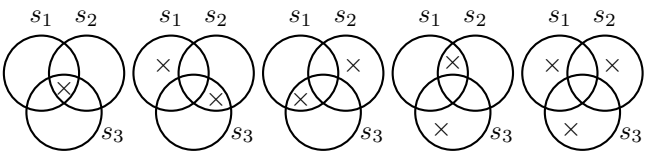
\begin{center}
\begin{minipage}{1.6cm}
\begin{center}
\begin{tikzpicture}[thick,scale=0.5]
        \draw (0,0) circle (1) node[above,shift={(0,0.5)}] {$s_1$};
        \draw (1.2,0) circle (1) node[above,shift={(0,0.5)}] {$s_2$};
        \draw (.6,-1.04) circle (1) node[shift={(0.65,-.3)}] {$s_3$};

        \node at (.6,-.4) {$\times$}; 
        \node at (1.2,-.7) {}; 
        \node at (0,-.7) {}; 
        \node at (1.4,.2) {}; 
        \node at (.6,.3) {}; 
        \node at (-.2,.2) {}; 
        \node at (.3,-1.5) {}; 
\end{tikzpicture}
\end{center}
\end{minipage}
\begin{minipage}{1.6cm}
\begin{center}
\begin{tikzpicture}[thick,,scale=0.5]
        \draw (0,0) circle (1) node[above,shift={(0,0.5)}] {$s_1$};
        \draw (1.2,0) circle (1) node[above,shift={(0,0.5)}] {$s_2$};
        \draw (.6,-1.04) circle (1) node[shift={(0.65,-.3)}] {$s_3$};

        \node at (.6,-.4) {}; 
        \node at (1.2,-.7) {$\times$}; 
        \node at (0,-.7) {}; 
        \node at (1.4,.2) {}; 
        \node at (.6,.3) {}; 
        \node at (-.2,.2) {$\times$}; 
        \node at (.3,-1.5) {}; 
\end{tikzpicture}
\end{center}
\end{minipage}
\begin{minipage}{1.6cm}
\begin{center}
\begin{tikzpicture}[thick,scale=0.5]
        \draw (0,0) circle (1) node[above,shift={(0,0.5)}] {$s_1$};
        \draw (1.2,0) circle (1) node[above,shift={(0,0.5)}] {$s_2$};
        \draw (.6,-1.04) circle (1) node[shift={(0.65,-.3)}] {$s_3$};

        \node at (.6,-.4) {}; 
        \node at (1.2,-.7) {}; 
        \node at (0,-.7) {$\times$}; 
        \node at (1.4,.2) {$\times$}; 
        \node at (.6,.3) {}; 
        \node at (-.2,.2) {}; 
        \node at (.3,-1.5) {}; 
\end{tikzpicture}
\end{center}
\end{minipage}
\begin{minipage}{1.6cm}
\begin{center}
\begin{tikzpicture}[thick,scale=0.5]
        \draw (0,0) circle (1) node[above,shift={(0,0.5)}] {$s_1$};
        \draw (1.2,0) circle (1) node[above,shift={(0,0.5)}] {$s_2$};
        \draw (.6,-1.04) circle (1) node[shift={(0.65,-.3)}] {$s_3$};

        \node at (.6,-.4) {}; 
        \node at (1.2,-.7) {}; 
        \node at (0,-.7) {}; 
        \node at (1.4,.2) {}; 
        \node at (.6,.3) {$\times$}; 
        \node at (-.2,.2) {}; 
        \node at (.3,-1.5) {$\times$}; 
\end{tikzpicture}
\end{center}
\end{minipage}
\begin{minipage}{1.6cm}
\begin{center}
\begin{tikzpicture}[thick,scale=0.5]
        \draw (0,0) circle (1) node[above,shift={(0,0.5)}] {$s_1$};
        \draw (1.2,0) circle (1) node[above,shift={(0,0.5)}] {$s_2$};
        \draw (.6,-1.04) circle (1) node[shift={(0.65,-.3)}] {$s_3$};

        \node at (.6,-.4) {}; 
        \node at (1.2,-.7) {}; 
        \node at (0,-.7) {}; 
        \node at (1.4,.2) {$\times$}; 
        \node at (.6,.3) {}; 
        \node at (-.2,.2) {$\times$}; 
        \node at (.3,-1.5) {$\times$}; 
\end{tikzpicture}
\end{center}
\end{minipage}
\end{center}
Alone, no sensor can distinguish between these 5 situations. 
To disambiguate between configurations, we must indeed use an inclusion-exclusion principle, so we need to identify if the same target has been detected by different sensors. For this, we suppose that, jointly, groups of sensors can determine if a target they saw was seen by all in that group. 
In this example, $s_1$ and $s_2$ together can for instance distinguish the worlds $w_1$ and $w_2$, but not $w_1$ and $w_4$. Only the three sensors together can distinguish between all five worlds. 
The framework we propose in this paper allows to formalize such applications. 


We show the very important, and surprising fact that every semi-simplicial set model is bisimilar to a simplicial complex model (leading to Theorem \ref{thm:complete-standard}).
There are several reasons why, despite this fact, we still believe that semi-simplicial set models are worth studying.
Indeed, some epistemic situations are described much more naturally and concisely using a semi-simplicial set model.
This is for instance the case of the sensor model described previously.
By turning it into a bisimilar simplicial complex model, we create a tree-like model with an infinite number of worlds, and where the underlying topology has disappeared.
For some applications (e.g.\ model-checking), it might be crucial to keep a model which is as small as possible, let alone infinite.
This situation is akin to classical topology: semi-simplicial sets and simplicial complexes describe the same spaces up to weak-homotopy. Yet, both structures are used in algebraic topology in their own right, according to the situation at hand. For some spaces, e.g.\ when dealing with the classification of combinatorial structures (posets, graphs, etc.), the simplicial complex approach is indeed very practical. However, in other situations (e.g.\ for non-triangulable spaces), the semi-simplicial set approach is much more convenient to use.

\paragraph*{Plan of the paper}

In \cref{sec:prelim}, we recall various mathematical notions that we will be using in this paper.
In \cref{sec:genKripke}, we define our two notions of models, generalized epistemic models and epistemic covering models; and prove that they are equivalent. 
We define various interesting sub-classes of these models in \cref{sec:subclasses}. In \cref{sec:axiomatization}, we prove that $\ECn$ is sound and complete with respect to these models (Theorem \ref{thm:complete-simple}), and give  sound and complete axiomatizations of interesting sub-classes (Theorems \ref{thm:complete-max}, \ref{thm:complete-min}, and \ref{thm:complete-pure}). Finally, we prove that every simplicial set model is bisimulation equivalent to a simplicial complex model, bringing interesting questions about the topological nature of bisimulation (Theorem \ref{thm:complete-standard}).

\section{Preliminaries}
\label{sec:prelim}

\subsection{Kripke semantics of distributed knowledge}
\label{sec:lang}
\label{sec:kripke-s5}

Let $A$ be a finite set of agents, and $\AP$ a countable set of atomic propositions.
We consider the language $\mathcal{L}_D$ of epistemic logic with \emph{distributed knowledge}~\cite{fagin}, generated by the following BNF grammar:
\[
\varphi ::= p \mid \neg\varphi \mid \varphi \land \varphi \mid
D_U\,\varphi \qquad p \in \AP,\ U \subseteq A
\]

The derived operators $\lor, \Rightarrow, \true, \false$, are defined as usual in propositional logic. We also use the following operators:
\begin{mathpar}
K_a\,\phi := D_{\{a\}}\,\phi \and
\widehat{D}_U\varphi := \neg D_U\neg \varphi
\end{mathpar}
The operator $D_U\phi$ is read ``the group of agents~$U$ (collectively) knows~$\phi$'', while its dual $\widehat{D}_U\varphi$ means that the group~$U$ considers possible the formula~$\phi$.
$K_a\,\phi$ is the standard knowledge operator of epistemic logic, ``agent $a$ knows~$\phi$''.

\begin{definition}
    A Kripke model is a structure $\mathcal{M} = {\la M,(\sim_a)_{a \in A}, L \ra}$, where:
    \begin{itemize}
        \item $M$ is a set of \emph{possible worlds},
        \item For every agent $a \in A$, $\sim_a$ is an equivalence relation,
        \item $L : M \to \mathscr{P}(\AP)$ is a valuation function.
    \end{itemize}
\end{definition}

Given a Kripke model, the satisfaction relation $M,w \models \phi$ is defined inductively as follows. We write $\sim_U \;=\; \bigcap_{a \in U} \sim_a$.
\[
\begin{array}{lcl}
    M,w \models p & \text{iff} & p \in L(w) \\
    M,w \models \neg\phi & \text{iff} & M,w \not\models \phi \\
    M,w \models \phi\land\psi & \text{iff} & M,w \models \phi \text{ and } M,w \models \psi \\
    M,w \models D_U\,\phi & \text{iff} & M,w' \models \phi \text{ for all } w' \in M \\ & & \text{ such that } w \sim_U w' 
\end{array}
\]

\subsection{Partial Equivalence Relations}

\begin{definition}
A \emph{Partial Equivalence Relation} (PER) on a set $X$ is a relation ${R \subseteq X \times X}$ which is symmetric and transitive (but not necessarily reflexive).
\end{definition}

The \emph{domain} of a PER $R$ is the set $\dom{R} = \{ x \in X \mid R(x,x)\} \subseteq X$, and it is easy to see that $R$ is an equivalence relation on its domain, and empty outside of it. Thus, a PER on~$X$ is simply an equivalence relation on a subset of~$X$.
The \emph{equivalence classes} of a PER~$R$ are defined as usual, when viewed as an equivalence relation on $\dom{R}$: for $x \in \dom{R}$, we write $[x]_R = \{ y \in X \mid x R y \}$ for the equivalence class of~$x$, and $X/R$ for the set of equivalence classes.

\subsection{Semi-simplicial sets}

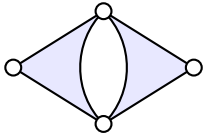
\begin{figure}[h]
    \centering
    \begin{tikzpicture}[scale=1.2]
	\begin{pgfonlayer}{nodelayer}
		\node [style=white vertex] (7) at (0, 1.25) {};
		\node [style=white vertex] (8) at (0, -1.25) {};
		\node [style=white vertex] (9) at (2, 0) {};
		\node [style=white vertex] (10) at (-2, 0) {};
	\end{pgfonlayer}
	\begin{pgfonlayer}{edgelayer}
		\draw [style=filled path] (7.center)
			 to (9.center)
			 to (8.center)
			 to [bend right=45] cycle;
		\draw [style=filled path] (8.center)
			 to (10.center)
			 to (7.center)
			 to [bend right=45] cycle;
	\end{pgfonlayer}
\end{tikzpicture}
    \caption{A simplicial set which is not a simplicial complex.}
    \label{fig:sset-ex}
\end{figure}

(Semi)-simplicial sets can be viewed as a higher-dimensional generalization of directed multigraphs, while simplicial complexes are a generalization of simple graphs.
Thus, in dimension~$1$, simplicial sets allow loops and parallel edges; whereas simplicial complexes do not. Similarly, in higher dimensions, simplicial sets allow simplices to be assembled together in more complex ways, as in \cref{fig:sset-ex}.
Standard references for the theory of simplicial sets and some of their many uses in algebraic topology are~\cite{goerss2009simplicial,may1993simplicial} and a more elementary introduction can be found in~\cite{introSimSetsFried2021}.

Compared to simplicial complexes, simplicial sets exhibit interesting geometric and categorical features:
\begin{itemize}
\item In algebraic topology, they provide a combinatorial model for the homotopy theory of topological spaces, in fact they are Quillen equivalent to the standard Quillen model category of topological spaces \cite{goerss2009simplicial}. 
\item As a presheaf category, simplicial sets form a Grothendieck topos. Still, simplicial complexes are almost as nice, they are known as forming a quasitopos, see e.g. \cite{Baez} for a discussion of the relationship between the quasitopos of simplicial complexes and the topos of simplicial sets.
\end{itemize}

Let $\Dpinj$ be the category of possibly empty linear orders with injective maps between them.
We write $[n]$ for the $(n+1)$-element linear order $[n] = \{0 < \ldots < n\}$, and $[-1]=\varnothing$.
The category of \emph{augmented semi-simplicial sets} is defined as the presheaf category $\widehat{\Delta}^+_{\inj}$.
Thus, its objects are functors $F : (\Dpinj)^{\op} \rightarrow \Set$, and morphisms are natural transformations.
The elements of $F([n])$ are called $n$-simplices, and we use the terms vertices, edges, and triangles for simplices of dimension~$0$, $1$ and $2$ respectively.

For example, the simplicial set depicted in \cref{fig:sset-ex} has $4$ vertices, $6$ edges and $2$ triangles. So, accordingly, the sets $F([0])$, $F([1])$ and $F([2])$ are chosen to have $4$, $6$ and $2$ elements, respectively.
There are three face maps $\partial_i : F([2]) \to F([1])$ for $i=0,1,2$ which assign to each triangle its three boundary edges. 
The maps $\partial_i : F([1]) \to F([0])$ for $i=0,1$ assign to each edge its source and target vertices.

\section{Generalized epistemic frames and epistemic coverings}
\label{sec:genKripke}

In this section, we define two classes of models for distributed knowledge: one generalizes Kripke frames, and the other generalizes simplicial models.
We show that they are structurally equivalent.
In the following sections, we will show that they subsume previous models found in the literature, and study their axiomatization.

\subsection{Generalized epistemic frames}

As in \cite{GoubaultLR21kb4} we will consider Kripke frames where the accessibility relation $\sim$ is not an equivalence relation but just a partial equivalence relation.
But we generalize one step further by associating an accessibility relation not only for each agent, but also for any set of agents, making it possible to interpret distributed knowledge in a very general manner.
Such models have been considered before in the epistemic logic literature, sometimes called ``pseudo-models’’, as an intermediate tool in completeness proofs involving distributed knowledge~\cite{Gerbrandy98distributed, BaltagS20}.
One paper that used such models as the main object of study is \cite{baltagCorrelatedKnowledgeEpistemicLogic2010}, in order to model observability in quantum systems.
Still, the models that we present here generalize further, by allowing worlds where not all agents are necessarily present, and as a consequence condition (b) below is new.

\begin{definition}
    \label{def:gef}
    A \emph{generalized epistemic frame} is a structure $\mathcal{M} = \la M,\sim \ra$, where:
    \begin{itemize}
        \item $M = \{w_0,w_1,\ldots\}$ is a set of \emph{possible worlds},
        \item $\sim$ is a function assigning for every group of agents $U \subseteq A$, a PER ${\sim_U \;\subseteq M \times M}$ called the \emph{$U$-accessibility relation}. 
        We write $[w]_{U}$ for the equivalence class of $w$ with respect to $\sim_U$.
        \item The PERs $\sim_U$ satisfy the following conditions.
        \begin{enumerate}[(a)]
        \item Compatibility:
        \begin{align*}
            \forall U' \subseteq U, \ & w \sim_U w' \Rightarrow w \sim_{U'} w' 
        \end{align*}
        \item Closure under union of the groups of alive agents:
        \begin{align*}
            \forall U, U' \subseteq A, \ & (w \sim_U w \land w\sim_{U'}w)\Rightarrow w \sim_{U \cup U'} w.
        \end{align*}
        \end{enumerate}
    \end{itemize}
\end{definition}

The two conditions on $\sim_U$ can be interpreted as follows.
Condition (a) means that if a group of agents cannot distinguish between two worlds, all together, then there is no way a subgroup of agents can distinguish the same two worlds.
Condition (b) implies that in each world $w$, there is a maximal group of agents $U$ such that $w \sim_U w$.
We call such a $U$ the group of \emph{alive agents} in world~$w$, and denote it by $\live{w}$.
We say that an agent~$a$ is \emph{alive} in~$w$ when $a \in \live{w}$ or, equivalently, when $w \sim_{\{a\}} w$.

Usually, we write $w \sim_{a} w'$ as shorthand for $w \sim_{\{a\}} w'$.
Additionally, we say that a world $w$ is a \emph{sub-world} of $w'$ when $\live{w} \subseteq \live{w'}$ and $w \sim_{\live{w}} w'$. 
The sub-world relation is a preorder, and is denoted by $w\leq w'$.

\begin{figure}[h]
    \centering
    \begin{tikzpicture}[auto, scale=1.4,font={\small}, {every loop/.style}={looseness=4, min distance=4mm}]
	\begin{pgfonlayer}{nodelayer}
		\node [style=empty node] (0) at (-1, 0) {$w_2$};
		\node [style=small font vertex] (2) at (1, 1) {$\{a,b,c\}$};
		\node [style=empty node] (5) at (1, 0) {$w_3$};
		\node [style=small font vertex] (7) at (0, 0.75) {$b$};
		\node [style=small font vertex] (9) at (-1, 1) {$\{a,b,c\}$};
		\node [style=small font vertex] (10) at (0, -0.75) {$c$};
		\node [style=empty node] (11) at (-3, 1) {$w_1$};
		\node [style=empty node] (12) at (3, 0) {$w_4$};
		\node [style=empty node] (13) at (5, 0) {$w_5$};
		\node [style=empty node] (14) at (-3, -1) {$w'_1$};
		\node [style=empty node] (15) at (-5, 0) {$w_0$};
		\node [style=small font vertex] (16) at (-3, 2) {$\{a,b\}$};
		\node [style=small font vertex] (17) at (-3, -2) {$\{a,b\}$};
		\node [style=small font vertex] (18) at (-5, 1) {$b$};
		\node [style=small font vertex] (19) at (-4, 0.75) {$b$};
		\node [style=small font vertex] (20) at (-4, -0.75) {$b$};
		\node [style=small font vertex] (21) at (-3.5, 0) {$\{a,b\}$};
		\node [style=small font vertex] (22) at (2, 0.25) {$\emptyset$};
		\node [style=small font vertex] (23) at (3, 1) {$\{a,c\}$};
		\node [style=small font vertex] (24) at (-2, 0.75) {$a$};
		\node [style=small font vertex] (25) at (-2, -0.75) {$a$};
	\end{pgfonlayer}
	\begin{pgfonlayer}{edgelayer}
		\draw [in=120, out=60, loop] (0) to ();
		\draw [bend left=330] (5) to (0);
		\draw [in=60, out=120, loop] (5) to ();
		\draw [bend right] (0) to (5);
		\draw (11) to (14);
		\draw (14) to (0);
		\draw (11) to (0);
		\draw (11) to (15);
		\draw (15) to (14);
		\draw [in=120, out=60, loop] (11) to ();
		\draw [in=-120, out=-60, loop] (14) to ();
		\draw [in=120, out=60, loop] (15) to ();
		\draw (5) to (12);
		\draw [in=120, out=60, loop] (12) to ();
	\end{pgfonlayer}
\end{tikzpicture}
    \caption{An example of a generalized epistemic frame.}
    \label{fig:gef-ex}
    \end{figure}
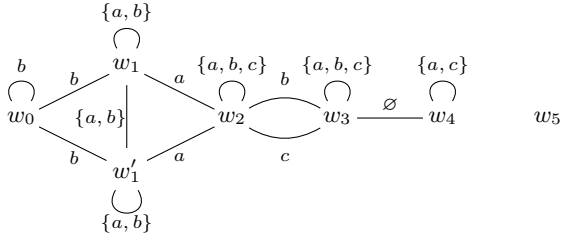

\begin{example}\label{ex:frame}
    An example of a generalized epistemic frame is given on \Cref{fig:gef-ex}.
    It has seven worlds $M = \{w_0, \ldots, w_5\} \cup \{w_1'\}$. Not all relations are shown, but only the generating ones.
    Sets of alive agents can be read off directly on the reflexive loops above a world; in~$w_5$, no agent is alive, that is $w_5\not\sim_a w_5$ for all~$a \in A$.
    The empty group can distinguish $w_5$ from other worlds: $w_5\not\sim_\emptyset w_i, i\not=5$.
    In world $w_4$, agents~$a$ and~$c$ are alive, and they can distinguish it from all other worlds, but the empty group cannot.
    In $w_2$ and $w_3$ all three agents are alive.
    Agents $b$ and~$c$ cannot individually distinguish $w_2$ from $w_3$, however, together they can: we have $w_2\sim_b w_3$ and $w_2\sim_c w_3$, but $w_2\not\sim_{\{b,c\}} w_3$.
    In $w_1$ and $w'_1$, agents~$a$ and~$b$ are alive, but even together they cannot distinguish them, as we have $w_1\sim_{\{a,b\}} w'_1$.
    World $w_0$ is a sub-world of both $w_1$ and $w'_1$: only~$b$ is alive in $w_0$ and it cannot distinguish $w_0$ from $w_1$ or~$w'_1$.

\end{example}

\begin{example}
\label{ex:sensnet}
In the example of the introduction, 
$w_1$, $\ldots$, $w_5$ are worlds, $s_1$, $s_2$, $s_3$ are agents, representing the sensors. For all $i,j,k$, we have $w_i \sim_{s_j} w_k$: no individual sensor $s_j$ can distinguish between any pair of worlds $w_i$ and $w_k$.
We also have $w_1 \sim_{\{s_2,s_3\}} w_2$ since in these two worlds, $s_2$ and $s_3$ jointly see the same target. More generally, 
$
w_1 \sim_{\{s_i,s_j\}} w_k 
$
if and only if some target is in $A_i \cap A_j$ in world $w_k$. 
But the three sensors together can distinguish $w_1$ from all other worlds. Examining all other intersections in pairs of worlds gives the  epistemic frame on \Cref{fig:sens-gef-ex}.

\begin{figure}[h]
    \centering

\newcommand\size{2}

\begin{tikzpicture}[every loop/.style={looseness=4, min distance=4mm}, every node/.style={inner sep=2pt}, font={\scriptsize}]
  \node[style=empty node] (w1) at (162:\size){$w_1$};
  \node[style=empty node] (w2) at (90:\size){$w_2$};
  \node[style=empty node] (w3) at (18:\size){$w_3$};
  \node[style=empty node] (w5) at (234:\size){$w_5$};
  \node[style=empty node] (w4) at (306:\size){$w_4$};
  \path (w1) edge[loop left] node {$\{1,2,3\}$} (w1)
              (w1) edge node[fill=white] {$1,\{2,3\}$} (w2)
              (w2) edge[loop above] node {$\{1,2,3\}$} (w2)
              (w2) edge node[fill=white] {$\{1,2\},3$} (w3)
              (w3) edge[loop right] node {$\{1,2,3\}$} (w3)
              (w3) edge node[fill=white] {$1,\{2,3\}$} (w4)
              (w4) edge[loop right] node {$\{1,2,3\}$} (w4)
              (w4) edge node[fill=white] {$\{1,3\},\{2,3\}$} (w5)
              (w5) edge[loop left] node {$\{1,2,3\}$} (w5)
              (w5) edge node[fill=white] {$1,2,3$} (w1)
              (w2) edge node[fill=white,pos=0.4] {$\{1,2\},\{1,3\}$} (w5)
              (w2) edge node[fill=white,pos=0.48] {$2,\{1,3\}$} (w4)
              (w3) edge node[sloped, fill=white, pos=0.4] {$\{1,2\},\{2,3\}$} (w5)
              (w1) edge node[sloped, fill=white,pos=0.4] {$\{1,2\},3$} (w4)
              (w1) edge node[above,fill=white] {$2,\{1,3\}$} (w3);
\end{tikzpicture}

    \caption{Sensor network as an epistemic frame.}
    \label{fig:sens-gef-ex}
    \end{figure}
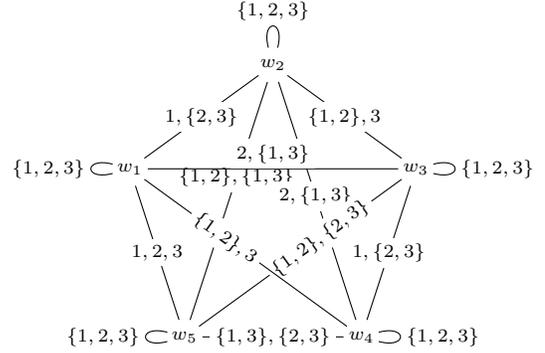
\end{example}

Morphisms between epistemic frames are structure-preserving functions between the sets of worlds.

\begin{definition}
    \label{def:gefmor}
    Let $\mathcal{M}=\langle M,\sim\rangle$ and $\mathcal{N}=\langle N, \sim'\rangle$ be two generalized epistemic frames. A morphism from $\mathcal{M}$ to $\mathcal{N}$ is a function $f: M \rightarrow N$ such that for all $U \subseteq A$, for all $u,v \in M$, $u \sim_U v$ implies $f(u) \sim'_U f(v)$, 
\end{definition}

We write $\GEF$ for the category of generalized epistemic frames with agents~$A$.
Later, in order to define the semantics of $\mathcal{L}_D$ formulas, we will equip these frames with a valuation function, as in \cref{sec:lang}.
But first, let us first define the geometric counterpart of these frames: epistemic coverings.

\subsection{Epistemic coverings}

\subsubsection{Chromatic augmented semi-simplicial sets}\label{sec:csets}

As in the case of simplicial models~\cite{gandalf-journal}, our first step will be to decorate the vertices of a simplicial set with colors, representing the names of the agents in~$A$.
The resulting structure is called a \emph{chromatic} augmented semi-simplicial set, or \emph{cset} for short. 
We identify $A$ with the linear order $[n]=\{0 < \ldots < n\}$. 

Let $S_A$ denote the \emph{standard $(|A|-1)$-simplex}, defined by:
\begin{itemize}
\item $(S_A)_k=\{({i_0},\ldots,{i_k}) \mid 0\leq i_0 < \ldots < i_k \leq n\}$,
\item $\partial_j({i_0},\ldots,{i_k})=(i_0,\ldots,{i_{j-1}},{i_{j+1}},\ldots,i_k) \in (S_A)_{k-1}$.
\end{itemize}
Given an augmented semi-simplicial set~$X$, a coloring of~$X$ by the agents in~$A$ is simply a map $f : X \to S_A$. 
Note that, since we work with \emph{semi}-simplicial sets here, without degeneracy maps, morphisms preserve the dimension of simplices. Thus, each simplex of~$X$ is well-colored, in the sense that all vertices in a simplex are labelled with distinct agents. 
We then define the category of chromatic augmented semi-simplicial sets to be the slice category $\widehat{\Delta}^+_{\inj}/S_A$. 

For the rest of the paper, we can either see this category as a slice category, 
or notice that, by the fundamental theorem of topos theory, 
the category of csets is once again a presheaf category 
on a site $\Gamma$ made of simplices of the standard $n$-simplex.
As with the site of semi-simplicial sets (see e.g., \cite{ss,Riehl1}), 
$\Gamma$ is the posetal category of subsets of $A$ 
with the inclusion partial order, defined below.

\begin{definition}
    \label{def:gamma}
    The category $\Gamma$ is such that:
    \begin{itemize}
        \item Objects are (possibly empty) subsets of~$A$.
        \item There is a unique morphism $\delta_{U,V} : U \to V$ in~$\Gamma$ whenever ${U \subseteq V}$. Composition $\delta_{V,W} \circ \delta_{U,V} = \delta_{U,W}$ is given by the fact that ${U \subseteq V \subseteq W}$ implies $U \subseteq W$.
    \end{itemize}
\end{definition}

We write $\Simp$ for the presheaf category on $\Gamma$. This category is equivalent to $\widehat{\Delta}^+_{\inj}/S_A$, hence a cset can equivalently be viewed as a functor $F : \Gamma^\op \to \Set$.
Given a cset $F \in \Simp$, and a group of agents $U \subseteq A$, the elements of $F(U)$ are called the \emph{$U$-simplices}.
When there is no ambiguity, we write $\partial_{U,V} : F(V) \to F(U)$ for the boundary operator $F(\delta_{U,V})$.
For $x$ a $V$-simplex, $\partial_{U,V}(x)$ is called the $U$-face of $x$. 
If it is clear which $V$ is considered, we simply write $\partial_U(x)$.

Given $U \subseteq A$, the standard $U$-simplex $\Gamma[U]$, is defined as the representable presheaf $\Gamma(-,U)$, image of~$U$ by the Yoneda embedding $y: \Gamma \to \Simp$.

\begin{example}
\label{ex:cset-base}
A cset~$X$ is depicted in the figure below. We represent it as a simplicial set together with colors on the vertices.
Elements of $X_{-1}$ are depicted as dashed regions (and interpreted as generalized connected components). Elements of $X_0$ are depicted as vertices, $X_1$ as edges, $X_2$ as triangles, etc. The boundary operators $\partial_i$ give the equations that permit to glue these simplices together, along lower dimensional simplices.
The cset in the picture below is composed of seven vertices (colored with three agents), eight edges, two triangles, and two $(-1)$-simplices (the annotations will only be used later).
%

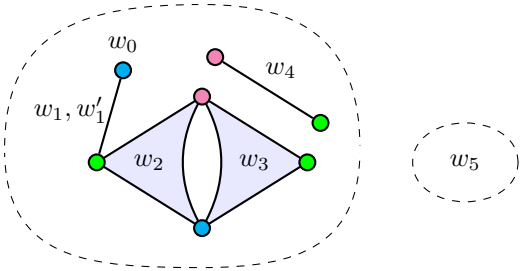
\begin{center}
    \begin{tikzpicture}[scale=1.4]
	\begin{pgfonlayer}{nodelayer}
		\node [style=green vertex] (3) at (2, 0) {};
		\node [style=cyan vertex] (4) at (-1.5, 1.75) {};
		\node [style=magenta vertex] (5) at (0.25, 2) {};
		\node [style=green vertex] (6) at (2.25, 0.75) {};
		\node [style=magenta vertex] (0) at (0, 1.25) {};
		\node [style=cyan vertex] (1) at (0, -1.25) {};
		\node [style=green vertex] (2) at (-2, 0) {};
		\node [style=empty node] (8) at (0, 3) {};
		\node [style=empty node] (9) at (-3.75, 0.25) {};
		\node [style=empty node] (10) at (0, -2) {};
		\node [style=empty node] (11) at (3, 0.25) {};
		\node [style=none] (12) at (4, 0) {};
		\node [style=none] (13) at (5, 0.75) {};
		\node [style=none] (14) at (6, 0) {};
		\node [style=none] (15) at (5, -0.75) {};
		\node [style=none] (17) at (-1, 0) {$w_2$};
		\node [style=none] (18) at (1, 0) {$w_3$};
		\node [style=none] (19) at (-1.5, 2.25) {$w_0$};
		\node [style=none] (20) at (1.5, 1.75) {$w_4$};
		\node [style=none] (21) at (-2.5, 1) {$w_1,w'_1$};
		\node [style=none] (22) at (5, 0) {$w_5$};
	\end{pgfonlayer}
	\begin{pgfonlayer}{edgelayer}
		\draw [style=thick edge] (5) to (6);
		\draw [style=thick edge] (4) to (2);
		\draw [style=filled path] (0.center)
			 to (3.center)
			 to (1.center)
			 to [bend right] cycle;
		\draw [style=filled path] (0.center)
			 to (2.center)
			 to (1.center)
			 to [bend left] cycle;
		\draw [style=dashed] (9.center)
			 to [in=-180, out=-90, looseness=1.25] (10.center)
			 to [in=-90, out=0, looseness=1.25] (11.center)
			 to [in=0, out=90, looseness=1.25] (8.center)
			 to [in=90, out=-180, looseness=1.25] cycle;
		\draw [style=dashed] (14.center)
			 to [in=0, out=90] (13.center)
			 to [in=90, out=-180] (12.center)
			 to [in=-180, out=-90] (15.center)
			 to [in=-90, out=0] cycle;
	\end{pgfonlayer}
\end{tikzpicture}
\end{center}
\end{example}

In the rest of the paper, by a \emph{(chromatic) simplicial complex} we mean a cset in which if two simplices share the same set of vertices, then they coincide.
More formally, if for two $U$-simplices $s,s'$ $\partial_a(s) = \partial_a(s')$ for all $a\in U$, then $s=s'$.

\subsubsection{Epistemic coverings}


In previous papers about simplicial models for epistemic logic, two issues have arisen:
\begin{enumerate}[(i)]
\item We need a way to describe which simplices of the model represent actual worlds, and which ones do not. Two canonical choices are possible: take all simplices as worlds, or take only the facets.
But one may also want to consider something in-between.
\item Simplicial models usually correspond to \emph{proper} Kripke models, because each individual simplex can represent only a single world.
\end{enumerate}
In this section, we solve both issues by introducing a new notion called \emph{epistemic coverings}.

The idea is that, lying above the simplicial set~$B$ representing the geometry of the model, we have a projective simplicial set~$E$ representing the worlds. A morphism $f: E \to B$ then maps each world to its representation in the geometric model.
This allows us to (i) freely assign worlds to the simplices of the model, and (ii) possibly assign several worlds to the same simplex.

In a cset $X$, given a $V$-simplex $t$ and a subset $U \subseteq V$, we say that $s = \partial_U(t)$ is a \emph{subsimplex} of $t$.
We say that~$X$ is \emph{projective} when for every simplex $s\in X$, there is a unique maximal simplex $\uparrow\!s$, 
such that $s$ is a subsimplex of $\uparrow\!s$.

\begin{definition}
    An \emph{epistemic covering} is a morphism $f:E\to B$ in $\Simp$ such that $E$ is projective and $f: E \to B$ is surjective.
    Equivalently, a morphism $f:E\to B$ is an epistemic covering if $E$ is projective and every maximal simplex of $B$ has a preimage.
    We refer to $E$ as \emph{top cset} and to $B$ as \emph{base cset}.
\end{definition}

Epistemic coverings form a category 
where morphisms from $f:E\to B$ to $f':E'\to B'$ are 
pairs of morphisms $\alpha_E:E\to E', \alpha_B:B\to B'$, 
such that the following square commutes: 
\begin{center}
\begin{tikzcd}
	E & {E'} \\
	B & {B'}
	\arrow["{\alpha_B}"', from=2-1, to=2-2]
	\arrow["f"', from=1-1, to=2-1]
	\arrow["{f'}", from=1-2, to=2-2]
	\arrow["{\alpha_E}", from=1-1, to=1-2]
\end{tikzcd}
\end{center}
We write this category $\eCov$. One can see that $\eCov$ is a full subcategory of the arrow category of $\Simp$.


An epistemic covering $f\!:\!E\to B$ can be visually represented as an annotated cset by taking the base cset $B$ and writing on a simplex $s$ the set of maximal simplices from $E$ that are mapped on $s$. In particular, as $f$ is surjective, every maximal simplex in $B$ must have an annotation. For instance, the annotation on the example of a cset from Example \ref{ex:cset-base} represents an epistemic covering with the following maximal simplices in $E$: two $2$-simplices $w_2, w_3$; three $1$-simplices $w_1, w'_1, w_4$ with $f(w_1)=f(w'_1)$; one $0$-simplex $w_0$; one $-1$-simplex $w_5$.

\begin{remark}\label{rem:simptosimp}
    There are two ways to canonically produce an epistemic covering out of a given cset~$X$.
    We can take $X$ as the base of the covering, but we need to choose the space~$E$.
    Choosing $E$ amounts to deciding which simplices of~$X$ constitute the possible worlds. 
    There are two natural choices: either take all of the simplices of~$X$, or take only the maximal ones (a.k.a.\ the facets).
    
    The first choice, that we call the ``maximal'' one, where $E$ is the disjoint union of all the simplices of $X$, yields a faithful functor from $\Simp$ to $\eCov$.
    A morphism $g: X\to Y$ is sent to the morphism of coverings 
    $\langle\alpha_E, \alpha_B \rangle$, 
    where $\alpha_B = g$ and $\alpha_E$ sends a simplex $s$ to $t$ if $g$ sends $s$ to $t$.
    This functor is injective on objects, so it makes $\Simp$ a subcategory of $\eCov$, though not full.
    The maximal interpretation appears (implicitly) in~\cite{Ditmarsch22Complete} for example, where a formula can be evaluated in every simplex of a model.
    
    The second choice, where $E$ is the disjoint union of the maximal simplices of~$X$, is the ``minimal'' one.
    It is the one that is studied in~\cite{gandalf-journal,GoubaultLR21kb4}, where a formula is only evaluated in a facet of a simplicial complex.
    However, this construction is not functorial.
\end{remark}

\subsection{Isomorphism between epistemic frames and coverings}

We want to show that the category of epistemic coverings 
and the category of generalized epistemic frames are isomorphic. 
We first define the functor $\kappa: \eCov\to \GEF$. 

Let $f: E\to B$ be an epistemic covering. 
As $E$ is projective, 
it can be represented as a disjoint union of disconnected standard simplices, 
that is $E=\bigcup_{U\subseteq A} \bigcup_{E_U} \Gamma[U]$, 
where $E_U$ are some sets. 
Then the set $W$ of worlds in $\kappa(f) = \langle W, \sim \rangle$ 
is exactly $\bigcup_{U\subseteq A} E_U$, 
that is every maximal simplex of $E$ is interpreted as a world. 
To define the group indistinguishability relations ${\sim_U}$ between the worlds of the corresponding frame, we proceed as follows. 
We think of two simplices in $E$ as $U$-indistinguishable,
if their images in $B$ share a $U$-face.
Formally, given two worlds $w_s, w_{s'}$ in $\kappa(f)$,
corresponding to two maximal simplices $s, s'$ in $E$,
we let $w_s\sim_U w_{s'}$ if $\partial_U(f(s))=\partial_U(f(s'))$.

Now we define this functor on morphisms. Suppose we are given a morphism of epistemic coverings $\langle \alpha_E, \alpha_B \rangle$. Then, the morphism of epistemic frames $\kappa(\langle \alpha_E, \alpha_B \rangle)\!:\! \kappa(f)\to \kappa(f')$ sends a world $w_s$ to a world $w_{s'}$ if the image of the maximal simplex $s$ in $E$ is included in the maximal simplex $s'$ in $E'$.

Next, we construct the inverse functor $\sigma: \GEF \to \eCov$.
Given a generalized frame $\langle M, \sim \rangle$, 
we need to define two csets $E_M$ and $B_M$ 
together with a map $m: E_M\to B_M$. 
For every world $w\in M$ 
we associate a simplex of type $\overline{w}$ to $E_M$, 
that is $E_M = \bigcup_{w\in M} \Gamma[\overline{w}]$.

It is slightly more intricate to build the base space~$B_M$ of the covering.
We construct it as a presheaf on~$\Gamma$, $B_M : \Gamma^\op \to \Set$.
For every group $U \subseteq A$ of agents, define $B_M(U) = M/_{\sim_U}$, the quotient of $M$ with respect to $\sim_U$.
The restriction $\partial_{U,V} : M/_{\sim_V} \to M/_{\sim_U}$ sends the equivalence class $[w]_V$ to $[w]_U$.
%
We need to verify that $B_M$ is indeed a cset.

\begin{lemma}
    For any frame $\langle M, \sim \rangle$, $B_M$ is a cset.
\end{lemma}
\begin{IEEEproof}
    First, let us show that restriction maps are correctly defined.
    Consider $\partial_{U,V}$ for $U\subseteq V$, and let $w' \in [w]_U$ be another representative of the equivalence class $[w]_U$.
    By monotonicity, we have 
    $w\sim_{V}w' \Rightarrow w\sim_{U}w'$, so they also belong to the same equivalence class $[w]_U = [w']_U$.
    Thus, the function~$\partial_{U,V}$ is correctly defined.
    To see that this is functorial, given sets of agents $U \subseteq V \subseteq W$, we need to prove that $\partial_{U,V} \circ \partial_{V,W} = \partial_{U,W}$, which is straightforward. 
    %
\end{IEEEproof}

One can see that 
there is a canonical map ${m : E_M \to B_M}$ 
which sends a maximal simplex $w$ of $E_M$,
representing a world,
to the corresponding equivalence class $[w]_{\live{w}} \in B_M(\live{w})$. This describes $\sigma$ on objects by setting $\sigma(\langle M, \sim \rangle) = m$.

To define $\sigma$ on morphisms, consider a morphism of epistemic frames $g: \mathcal{M}\to \mathcal{N}$. As both assignments of total spaces $E_M, E_N$ and base spaces $B_M, B_N$ are functorial one needs to check that the induced square commutes. 

\begin{proposition}\label{prop:main} 
    The functors $\kappa$ and $\sigma$ define an isomorphism of categories: $\kappa\circ \sigma = \id_{\GEF}$ and $\sigma\circ \kappa = \id_{\eCov}$.
\end{proposition}
\begin{IEEEproof}
    Consider an epistemic frame $\mathcal{M}$. Then $\kappa\sigma(\mathcal{M})$ has as its worlds the same worlds as in $\mathcal{M}$, as $\kappa$ and $\sigma$ just transfer this information. The relations $\sim_U$ are also just the same: $w\sim_U w'$ in $\kappa\sigma(\mathcal{M})$ iff the simplices $\partial_U(w),\partial_U(w')$ in $E_{\sigma(\mathcal{M})}$ are sent to the same simplex in $B_{\sigma(\mathcal{M})}$, but this is the case exactly when $w\sim_U w'$ in $\mathcal{M}$.
    The same line of argument works for $\sigma\circ \kappa$, and extends to morphisms.
\end{IEEEproof}

\begin{example}
Let us illustrate the isomorphism using on the epistemic frame~$\mathcal M$ of \Cref{ex:frame}. We now describe its equivalent representation as an epistemic covering~$\sigma(\mathcal M)$.
In fact, the base~$B$ of the covering is the one depicted in \Cref{ex:cset-base}.
Then, we need a projective cset~$E$, which can be thought of as an ``exploded view'' of~$B$, depicted below.
Its maximal simplices, labelled $\{w_0,\dots,w_5\} \cup \{w_1'\}$, correspond to the worlds of the original frame~$\mathcal M$.
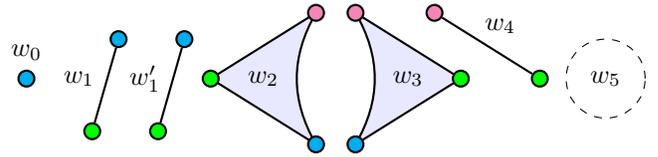
\begin{center}
	\smallskip
    \begin{tikzpicture}[scale=1.4]
	\begin{pgfonlayer}{nodelayer}
		\node [style=green vertex] (3) at (2, 1) {};
		\node [style=magenta vertex] (5) at (1.5, 2.25) {};
		\node [style=green vertex] (6) at (3.5, 1) {};
		\node [style=magenta vertex] (0) at (0, 2.25) {};
		\node [style=cyan vertex] (1) at (0, -0.25) {};
		\node [style=none] (12) at (4, 1) {};
		\node [style=none] (13) at (4.75, 1.75) {};
		\node [style=none] (14) at (5.5, 1) {};
		\node [style=none] (15) at (4.75, 0.25) {};
		\node [style=cyan vertex] (16) at (-4.5, 1.75) {};
		\node [style=green vertex] (17) at (-5, 0) {};
		\node [style=cyan vertex] (18) at (-3.25, 1.75) {};
		\node [style=green vertex] (19) at (-3.75, 0) {};
		\node [style=magenta vertex] (20) at (-0.75, 2.25) {};
		\node [style=cyan vertex] (21) at (-0.75, -0.25) {};
		\node [style=green vertex] (22) at (-2.75, 1) {};
		\node [style=none] (23) at (-5.25, 1) {};
		\node [style=none] (24) at (-5.25, 1) {$w_1$};
		\node [style=none] (25) at (-4, 1) {$w_1'$};
		\node [style=none] (26) at (-1.75, 1) {};
		\node [style=none] (27) at (-1.75, 1) {$w_2$};
		\node [style=none] (28) at (1, 1) {$w_3$};
		\node [style=none] (29) at (2.75, 2) {$w_4$};
		\node [style=none] (30) at (4.75, 1) {$w_5$};
		\node [style=cyan vertex] (31) at (-6.25, 1) {};
		\node [style=none] (32) at (-6.25, 1.5) {$w_0$};
	\end{pgfonlayer}
	\begin{pgfonlayer}{edgelayer}
		\draw [style=thick edge] (5) to (6);
		\draw [style=filled path] (0.center)
			 to (3.center)
			 to (1.center)
			 to [bend right] cycle;
		\draw [style=dashed] (14.center)
			 to [in=0, out=90] (13.center)
			 to [in=90, out=-180] (12.center)
			 to [in=-180, out=-90] (15.center)
			 to [in=-90, out=0] cycle;
		\draw [style=thick edge] (16) to (17);
		\draw [style=thick edge] (18) to (19);
		\draw [style=filled path] (20.center)
			 to (22.center)
			 to (21.center)
			 to [bend left] cycle;
	\end{pgfonlayer}
\end{tikzpicture}
    \smallskip
\end{center}
The covering $f : E \to B$ maps each world of $E$ to its geometric representation in the base~$B$.
Note that both $w_1$ and $w_1'$ are mapped to the same edge of~$B$: this is how we model non-proper behavior in epistemic frames.
Moreover, world~$w_0$ is represented by a single vertex, because only one agent is alive; it is mapped to the top blue vertex of~$B$. This is how we model sub-worlds: since $w_0$ is a sub-world of~$w_1$ in~$\mathcal M$, $f(w_0)$ is a sub-simplex of $f(w_1)$ in~$\sigma(\mathcal M)$.
%
\end{example}

\subsection{Semantics of distributed knowledge}
\label{sec:generalized-models-semantics}

We now use generalized epistemic frames, and equivalently epistemic coverings, as a model for the logic of distributed knowledge.
The missing piece of data is to label the worlds with \emph{atomic propositions}, in order to specify which facts about the system are either true or false in any given world.

\begin{definition}
    A \emph{(generalized) epistemic model} $\mathcal{M} = {\langle M, \sim, L \rangle}$ over the set of agents $A$ consists of a generalized frame $\langle M, \sim \rangle$ together with \emph{valuation} function $L: M \to \mathscr{P}(\AP)$.
    A morphism of epistemic models $f: \mathcal{M}\to \mathcal{N}$
    is a morphism of underlying frames that preserves valuations,
    that is, if $p\in L_M(w)$, then $p\in L_N(f(w))$.
    The category of generalized epistemic models is denoted $\mathcal{KM}_A$.
\end{definition} 

\begin{remark}
    Morphisms in $\mathcal{KM}_A$ are also known as functional simulations~\cite{blackburnModalLogic2001}.
    They are different from the morphisms used in \cite{gandalf-journal}:
    there the valuations of atomic propositions were preserved and reflected,
    that is $L_M(s) = L_N(f(s))$, whereas in our definition $L_M(s)\subseteq L_N(f(s))$. 
    They are also different from morphisms in \cite{GoubaultLR21kb4}:
    there, morphisms can be seen as relations, as they are maps $f: M\to \mathscr{P}(N)$.
\end{remark}





Given a generalized epistemic model, we can define the satisfaction relation as we did in \cref{sec:kripke-s5}.
Note however that $\sim_U$ is now part of the structure of the model, and might not be equal to $\bigcap_{a \in U} \sim_a$ in general.
\[
\begin{array}{lcl}
    M,w \models D_U\,\phi & \text{iff} & M,w' \models \phi \text{ for all } w' \in M \\ & & \text{ such that } w \sim_U w' 
\end{array}
\]
Similarly, we can equip epistemic coverings with a valuation:

\begin{definition}
An \emph{epistemic covering model} $X = \langle {f : E\to B}, \ell \rangle$
consists of an epistemic covering $f\!:\! E\to B$, together with
a labelling $\ell: \max(E)\to \mathscr{P}(\AP)$ that associates with each
maximal simplex $s$ of $E$ a set of atomic propositions~$\ell(s)$ that hold there. 
A morphism of epistemic covering models 
$\alpha : X \to Y$
is a morphism of epistemic coverings that preserves the labelling:
$\ell(s) \subseteq \ell'(\uparrow\!\alpha(s))$,
where $\uparrow\!\alpha(s)$ is the maximal simplex of $E$ that
contains $\alpha(s)$.
We denote by $\mathcal{EC}_A$ 
the category of epistemic covering models.
\end{definition}

Given an epistemic covering model $X = \langle f, \ell \rangle$ together with a maximal simplex $s$ in $E$,
we can define the satisfaction relation $X,s\models \varphi$
by transporting what we did for generalized epistemic models via the isomophism of Proposition \ref{prop:main}.
\[\begin{array}{lcl}
    X,s \models D_U\,\phi & \text{iff} & X,s' \models \phi \text{ for all maximal } s' \in E \\ & & \text{ such that } \partial_U(f(s')) = \partial_U(f(s))
\end{array}
\]
\Cref{prop:main} can be readily extended to show that the categories of models are also isomorphic:

\begin{theorem}\label{prop:main2}
    The category of epistemic covering models $\mathcal{EC}_A$ is
    isomorphic to the category of generalized epistemic models $\mathcal{KM}_A$.
\end{theorem}
\begin{IEEEproof}
    We provide two functors $\kappa,\sigma$. 
    On the underlying coverings and frames, they act as in \Cref*{prop:main}.
    We just need to extend them to valuations. 
    For $\kappa: \mathcal{EC}_A \to \mathcal{KM}_A$, 
    for each maximal simplex $s$ in $E$, we have an associated world $w_s$ in $\kappa(f)$. We set $L_{\kappa(f)}(w_s) = \ell(s)$.
    Similarly, for ${\sigma:\mathcal{KM}_A \to \mathcal{EC}_A}$,
    there is an associated maximal simplex $s_w$ for every world $w$.
    We set $\ell_{\sigma(M)}(s_w) = L(w)$. 
    The rest of the proof is the same as in \Cref{prop:main}.
\end{IEEEproof}

As expected, the satisfaction relations for both kinds of models yield the same result:

\begin{lemma}\label{lem:equiv-sat}
    Given a pointed epistemic covering model $(X,s)$,
    we have $X,s\models \varphi$ iff $\kappa(X), w_s\models \varphi$.
    Conversely, given a pointed generalized epistemic model $(M,w)$,
    we have $M,w\models \varphi$ iff $\sigma(M), s_w \models \varphi$.
\end{lemma}
\begin{IEEEproof}
    We prove the first equivalence by induction on the structure of~$\varphi$.
    The case of atomic propositions comes from the fact
    that we keep the labelling $L(w_s)=\ell(s)$.
    The case of boolean connectives is straightforward.
    For a formula of the form $D_U\varphi$
    one can notice that we defined $w_s\sim_U w_{s'}$
    iff $\partial_U(f(s))=\partial_U(f(s'))$,
    which coincides with the semantics of~$D_U$.
    The second equivalence follows from the first one, together with \cref{prop:main2}.
%
\end{IEEEproof}

\section{Properties of epistemic models}
\label{sec:subclasses}

The models presented in \cref{sec:generalized-models-semantics} are very versatile.
Depending on what kind of applications we have in mind, we might want to impose some extra properties on the structure of our models.
For instance in distributed computing, the model is usually assumed to be a simplicial complex rather than a simplicial set, because a global state of the system is merely the sum of the local states of the agents, without any extra information.
In this section, we define a number of interesting properties of epistemic frames, as well as their geometric counterpart, epistemic coverings.
We will see how some of the results of previous papers on simplicial models arise as a special case of  \Cref{prop:main}.

\subsection{Properties of epistemic frames and coverings}

\begin{definition}
\label{def:frame-properties}
    An epistemic frame $\mathcal{M} = \langle M, \sim \rangle$ is said~to
    \begin{itemize}
        \item have \emph{trivial empty-group knowledge} \\
        if $\forall w,w'\in M.\ w\sim_\varnothing w'$;
        \item have \emph{no empty worlds}
        if $\forall w\in M.\ \exists a\in A.\ w\sim_a w$;
        \item be \emph{proper}
        if $(\overline{w} = \overline{w'} \land w\sim_{\live{w}} w') \Rightarrow w = w'$;
        \item be \emph{maximal}
        if $\forall w\in M.\ \forall U\subseteq \live{w}.$ \\
        $U \neq \emptyset \Rightarrow \exists w'\in M.\ (U=\live{w'} \land w' \sim_{U} w)$;
        \item be \emph{minimal} 
        if $\forall w,w'\in M.\ (\live{w}\subsetneq \live{w'}) \Rightarrow w\not\sim_{\live{w}} w'$;
        \item be \emph{pure}
        if $\forall w\in M.\ \live{w} = A$; 
        \item have \emph{standard group knowledge} \\
        if $\forall U\!\subseteq\! A.\ (\forall a\in U.\ w\sim_a w') \Rightarrow w\sim_U w'$.
    \end{itemize}
\end{definition}

Let us explain the meaning of these properties.
In a frame with trivial empty-group knowledge,
the empty group cannot distinguish any worlds. 
It models the idea that an empty group cannot measure anything, so all worlds have the same properties for it. 

If a frame has no empty worlds,
then there is an alive agent in every world,
i.e., every possibility is observed by someone.

A frame is proper if every pair of worlds that 
has the same set of alive agents is distinguishable
by some subgroup of agents.
This also corresponds to the principle of observability:
if even a maximal group cannot distinguish worlds, then they are the same.
Notice that this allows sub-worlds.

\begin{example}
In the figure below, the leftmost frame has trivial empty-group knowledge, ($w_0\sim_{\emptyset}w_1\sim_{\emptyset}w_2$), but also has an empty world $w_2$, and is not proper ($w_0\sim_{\{a,b\}}w_1$).
The middle frame has non-trivial empty-group knowledge (${w_1\not\sim_\emptyset w_2}$), has no empty worlds, and is not proper.
The rightmost frame has non-trivial empty-group knowledge, has an empty world, but is proper ($w_0\not\sim_b w_1$).

	\hspace{-0.5cm}
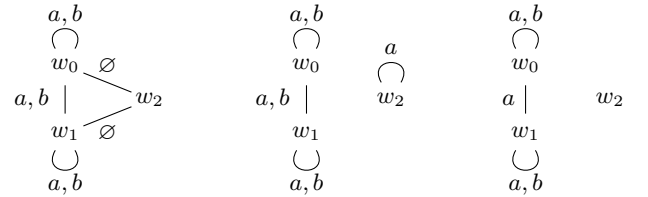
    \begin{tabular}{l@{\hskip 0.7cm}l@{\hskip 0.7cm}l}
        \begin{tikzpicture}[auto,  scale=0.9,font={\small}, {every loop/.style}={looseness=4, min distance=4mm}]
	\begin{pgfonlayer}{nodelayer}
		\node [style=empty node] (0) at (-1, 1) {$w_0$};
		\node [style=empty node] (1) at (1.5, 0) {$w_2$};
		\node [style=new style 0] (2) at (-1, 2.5) {$a,b$};
		\node [style=new style 0] (4) at (0.25, 1) {$\emptyset$};
		\node [style=empty node] (5) at (-1, -1) {$w_1$};
		\node [style=new style 0] (6) at (0.25, -1) {$\emptyset$};
		\node [style=new style 0] (7) at (-2, 0) {$a,b$};
		\node [style=new style 0] (8) at (-1, -2.5) {$a,b$};
	\end{pgfonlayer}
	\begin{pgfonlayer}{edgelayer}
		\draw (0) to (1);
		\draw [in=120, out=60, loop] (0) to ();
		\draw (5) to (0);
		\draw (5) to (1);
		\draw [in=-60, out=-120, loop] (5) to ();
	\end{pgfonlayer}
\end{tikzpicture} &
        \begin{tikzpicture}[auto, scale=0.9, font={\small}, {every loop/.style}={looseness=4, min distance=4mm}]
	\begin{pgfonlayer}{nodelayer}
		\node [style=empty node] (0) at (-1, 1) {$w_0$};
		\node [style=empty node] (1) at (1.5, 0) {$w_2$};
		\node [style=new style 0] (2) at (-1, 2.5) {$a,b$};
		\node [style=empty node] (5) at (-1, -1) {$w_1$};
		\node [style=new style 0] (7) at (-2, 0) {$a,b$};
		\node [style=new style 0] (8) at (-1, -2.5) {$a,b$};
		\node [style=new style 0] (9) at (1.5, 1.5) {$a$};
	\end{pgfonlayer}
	\begin{pgfonlayer}{edgelayer}
		\draw [in=120, out=60, loop] (0) to ();
		\draw (5) to (0);
		\draw [in=-60, out=-120, loop] (5) to ();
		\draw [in=120, out=60, loop] (1) to ();
	\end{pgfonlayer}
\end{tikzpicture} &
        \begin{tikzpicture}[auto,  scale=0.9,font={\small}, {every loop/.style}={looseness=4, min distance=4mm}]
	\begin{pgfonlayer}{nodelayer}
		\node [style=empty node] (0) at (-1, 1) {$w_0$};
		\node [style=empty node] (1) at (1.5, 0) {$w_2$};
		\node [style=new style 0] (2) at (-1, 2.5) {$a,b$};
		\node [style=empty node] (5) at (-1, -1) {$w_1$};
		\node [style=new style 0] (7) at (-1.5, 0) {$a$};
		\node [style=new style 0] (8) at (-1, -2.5) {$a,b$};
	\end{pgfonlayer}
	\begin{pgfonlayer}{edgelayer}
		\draw [in=120, out=60, loop] (0) to ();
		\draw (5) to (0);
		\draw [in=-60, out=-120, loop] (5) to ();
	\end{pgfonlayer}
\end{tikzpicture}
    \end{tabular}
\end{example}

A frame is maximal if every world has a non-empty sub-world. 
A certain intuition comes from distributed computing: in a maximal frame, any number of agents may crash during the execution of a program, as long as at least one of them remains alive. Moreover, these crashes are undetectable.

A frame is minimal if there are no strict sub-worlds. Once again, this corresponds to a situation in distributed computing where crashes are detectable, that is when
a process crashes, one of the remaining processes is aware of it.

A frame is pure if the set of alive agent in every world is the same. In such a situation, crashes are not allowed at all, and all agents always participate.

By ``standard group knowledge'', we mean that group indistinguishability relations are generated by individual agents,
that is, knowledge of the group is exactly 
the sum of individual agents' knowledge.
More formally, when $\sim_U \;=\; \bigcap_{a\in U} \sim_a$.

\begin{example}
In the figure below, the upper left frame is maximal, as it has all sub-worlds.
The upper right frame is minimal as it does not have sub-worlds at all.
The bottom left frame is pure as all its worlds have the same set of alive agents.
The bottom right frame has non-standard group knowledge, contrary to all previous examples, since $w_0\sim_a w_1$, $w_0\sim_b w_1$, but $w_0\not\sim_{a,b}w_1$.

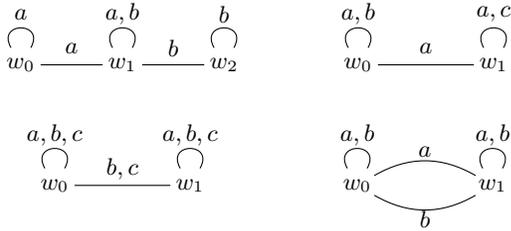
\begin{center}
    \begin{tabular}{c@{\hskip 1cm}c}
    \begin{tikzpicture}[auto, scale=0.9, font={\small}, {every loop/.style}={looseness=4, min distance=4mm}]

            \begin{pgfonlayer}{nodelayer}
                \node [style=empty node] (0) at (-1.5, 0) {$w_0$};
                \node [style=empty node] (1) at (1.5, 0) {$w_2$};
                \node [style=new style 0] (2) at (0, 0.75) {$a,b$};
                \node [style=empty node] (5) at (0, 0) {$w_1$};
                \node [style=new style 0] (7) at (-0.75, 0.25) {$a$};
                \node [style=new style 0] (9) at (-1.5, 0.75) {$a$};
                \node [style=new style 0] (10) at (1.5, 0.75) {$b$};
                \node [style=new style 0] (11) at (0.75, 0.25) {$b$};
            \end{pgfonlayer}
            \begin{pgfonlayer}{edgelayer}
                \draw [in=120, out=60, loop] (0) to ();
                \draw (5) to (0);
                \draw [in=60, out=120, loop] (5) to ();
                \draw [in=120, out=60, loop] (1) to ();
                \draw (5) to (1);
            \end{pgfonlayer}
        
    \end{tikzpicture}

    &

    \begin{tikzpicture}[auto, scale=0.9, font={\small}, {every loop/.style}={looseness=4, min distance=4mm}]
            \begin{pgfonlayer}{nodelayer}
                \node [style=empty node] (0) at (-1, 0) {$w_0$};
                \node [style=new style 0] (2) at (1, 0.75) {$a,c$};
                \node [style=empty node] (5) at (1, 0) {$w_1$};
                \node [style=new style 0] (7) at (0, 0.25) {$a$};
                \node [style=new style 0] (9) at (-1, 0.75) {$a,b$};
            \end{pgfonlayer}
            \begin{pgfonlayer}{edgelayer}
                \draw [in=120, out=60, loop] (0) to ();
                \draw (5) to (0);
                \draw [in=60, out=120, loop] (5) to ();
            \end{pgfonlayer}
        \end{tikzpicture}
        
    \\

    \begin{tikzpicture}[auto, scale=0.9, font={\small}, {every loop/.style}={looseness=4, min distance=4mm}, baseline={(0,0)}]
        \begin{pgfonlayer}{nodelayer}
            \node [style=empty node] (0) at (-1, 0) {$w_0$};
            \node [style=new style 0] (2) at (1, 0.75) {$a,b,c$};
            \node [style=empty node] (5) at (1, 0) {$w_1$};
            \node [style=new style 0] (7) at (0, 0.25) {$b,c$};
            \node [style=new style 0] (9) at (-1, 0.75) {$a,b,c$};
        \end{pgfonlayer}
        \begin{pgfonlayer}{edgelayer}
            \draw [in=120, out=60, loop] (0) to ();
            \draw (5) to (0);
            \draw [in=60, out=120, loop] (5) to ();
        \end{pgfonlayer}
    \end{tikzpicture}

    &

    \begin{tikzpicture}[auto, scale=0.9, font={\small}, {every loop/.style}={looseness=4, min distance=4mm}, baseline={(0,0)}]
            \begin{pgfonlayer}{nodelayer}
                \node [style=empty node] (0) at (-1, 0) {$w_0$};
                \node [style=new style 0] (2) at (1, 0.75) {$a,b$};
                \node [style=empty node] (5) at (1, 0) {$w_1$};
                \node [style=new style 0] (7) at (0, 0.5) {$a$};
                \node [style=new style 0] (9) at (-1, 0.75) {$a,b$};
                \node [style=new style 0] (10) at (0, -0.5) {$b$};
            \end{pgfonlayer}
            \begin{pgfonlayer}{edgelayer}
                \draw [in=120, out=60, loop] (0) to ();
                \draw [bend left=330] (5) to (0);
                \draw [in=60, out=120, loop] (5) to ();
                \draw [bend right] (0) to (5);
            \end{pgfonlayer}
        \end{tikzpicture}
    \end{tabular}
    \end{center}
\end{example}

Similarly, for epistemic coverings: 

\begin{definition}
\label{def:covering-properties}
    An epistemic covering $f: E\to B$ is said~to
    \begin{itemize}
        \item have \emph{trivial empty-group knowledge} 
        if there is only one simplex of dimension $-1$ in $B$;
        \item have \emph{no empty worlds} 
        if all maximal simplices of~$E$ have dimension~$\geq 0$;
        \item be \emph{proper} if
        no two maximal simplices of~$E$ have
        the same image in $B$;
        \item be \emph{maximal} if every simplex in $B$ is the image of a maximal simplex of~$E$;
        \item be \emph{minimal} if the image of a maximal simplex of~$E$ is always a maximal simplex of~$B$;
        \item be \emph{pure} if all maximal simplices of $E$ have dimension $|A| - 1$;
        \item have \emph{standard group knowledge} if
        $B$ is a simplicial complex.
    \end{itemize}
\end{definition}

The intuition behind the definitions is exactly the same as in the case of frames. We give a few illustrative examples. 
In the picture below, the leftmost covering has trivial empty-group knowledge as there is only one $(-1)$-simplex in the base space (one dashed region); but it has empty worlds because of the maximal $(-1)$-simplex $w_2$.
The covering depicted in the middle has no empty worlds since all the maximal simplices have dimension $0$ or $1$. It, however, does not have trivial empty-group knowledge as it has two $(-1)$-simplices in the base space (depicted as two dashed regions). It is not proper either because both worlds $w_0, w_1$ label the same edge.
The rightmost covering is proper, as every simplex is annotated with at most one world. It has an empty world, $w_2$, and does not have trivial empty-group knowledge.
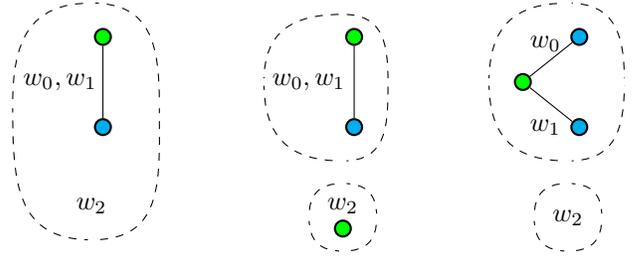
\begin{center}
\vspace{-8pt}
\begin{tabular}{c@{\hskip 1cm}c@{\hskip 1cm}c}
        \begin{tikzpicture}[auto, scale=1.2]
	\begin{pgfonlayer}{nodelayer}
		\node [style=cyan vertex] (2) at (-0.75, -1) {};
		\node [style=empty node] (8) at (-1, 1.75) {};
		\node [style=empty node] (9) at (-2.75, -0.5) {};
		\node [style=empty node] (10) at (-1, -3.5) {};
		\node [style=empty node] (11) at (0.5, -0.5) {};
		\node [style=none] (17) at (-1, -2.75) {$w_2$};
		\node [style=none] (18) at (-1.7, 0) {$w_0,w_1$};
		\node [style=green vertex] (19) at (-0.75, 1) {};
	\end{pgfonlayer}
	\begin{pgfonlayer}{edgelayer}
		\draw [style=dashed] (9.center)
			 to [in=-180, out=-90, looseness=1.25] (10.center)
			 to [in=-90, out=0, looseness=1.25] (11.center)
			 to [in=0, out=90, looseness=1.25] (8.center)
			 to [in=90, out=-180, looseness=1.25] cycle;
		\draw (2) to (19);
	\end{pgfonlayer}
\end{tikzpicture} &
        \begin{tikzpicture}[auto, scale=1.2]
	\begin{pgfonlayer}{nodelayer}
		\node [style=cyan vertex] (2) at (-0.75, -1) {};
		\node [style=empty node] (8) at (-1, 1.5) {};
		\node [style=empty node] (9) at (-2.75, 0) {};
		\node [style=empty node] (10) at (-1, -1.75) {};
		\node [style=empty node] (11) at (0, 0) {};
		\node [style=none] (17) at (-1, -2.75) {$w_2$};
		\node [style=none] (18) at (-1.75, 0) {$w_0,w_1$};
		\node [style=green vertex] (19) at (-0.75, 1) {};
		\node [style=green vertex] (20) at (-1, -3.25) {};
		\node [style=empty node] (21) at (-1, -2.25) {};
		\node [style=empty node] (22) at (-1.75, -3) {};
		\node [style=empty node] (23) at (-1, -3.75) {};
		\node [style=empty node] (24) at (-0.25, -3) {};
	\end{pgfonlayer}
	\begin{pgfonlayer}{edgelayer}
		\draw [style=dashed] (9.center)
			 to [in=-180, out=-90, looseness=1.25] (10.center)
			 to [in=-90, out=0, looseness=1.25] (11.center)
			 to [in=0, out=90, looseness=1.25] (8.center)
			 to [in=90, out=-180, looseness=1.25] cycle;
		\draw (2) to (19);
		\draw [style=dashed] (22.center)
			 to [in=-180, out=-90, looseness=1.25] (23.center)
			 to [in=-90, out=0, looseness=1.25] (24.center)
			 to [in=0, out=90, looseness=1.25] (21.center)
			 to [in=90, out=-180, looseness=1.25] cycle;
	\end{pgfonlayer}
\end{tikzpicture} &
   	    \begin{tikzpicture}[auto, scale=1.2]
	\begin{pgfonlayer}{nodelayer}
		\node [style=cyan vertex] (2) at (-0.75, -1) {};
		\node [style=empty node] (8) at (-1, 1.75) {};
		\node [style=empty node] (9) at (-2.75, 0) {};
		\node [style=empty node] (10) at (-1, -1.75) {};
		\node [style=empty node] (11) at (0.25, 0) {};
		\node [style=none] (17) at (-1, -3) {$w_2$};
		\node [style=none] (18) at (-1.5, 0.85) {$w_0$};
		\node [style=green vertex] (19) at (-2, 0) {};
		\node [style=empty node] (21) at (-1, -2.25) {};
		\node [style=empty node] (22) at (-1.75, -3) {};
		\node [style=empty node] (23) at (-1, -3.75) {};
		\node [style=empty node] (24) at (-0.25, -3) {};
		\node [style=cyan vertex] (25) at (-0.75, 1) {};
		\node [style=none] (26) at (-1.5, -1) {$w_1$};
	\end{pgfonlayer}
	\begin{pgfonlayer}{edgelayer}
		\draw [style=dashed] (9.center)
			 to [in=-180, out=-90, looseness=1.25] (10.center)
			 to [in=-90, out=0, looseness=1.25] (11.center)
			 to [in=0, out=90, looseness=1.25] (8.center)
			 to [in=90, out=-180, looseness=1.25] cycle;
		\draw (2) to (19);
		\draw [style=dashed] (22.center)
			 to [in=-180, out=-90, looseness=1.25] (23.center)
			 to [in=-90, out=0, looseness=1.25] (24.center)
			 to [in=0, out=90, looseness=1.25] (21.center)
			 to [in=90, out=-180, looseness=1.25] cycle;
		\draw (25) to (19);
	\end{pgfonlayer}
\end{tikzpicture}
\end{tabular}
\end{center}

In the examples below, the top left covering is maximal as every simplex is annotated, that is, every simplex has a maximal simplex that is sent to it. The top right covering is minimal because only maximal simplices are annotated.
The bottom left covering is pure, as all annotated simplices are of the same dimension. All of the examples above have standard group knowledge since their base csets are in fact complexes. The bottom right covering has non-standard group knowledge as its base cset is not a simplicial complex.

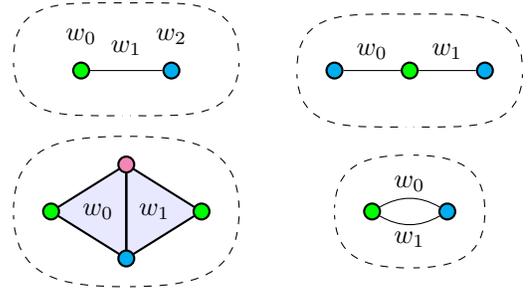
\begin{center}
    \begin{tabular}{cc}
    \begin{tikzpicture}[auto, scale=1.2]
	\begin{pgfonlayer}{nodelayer}
		\node [style=cyan vertex] (2) at (1, 0) {};
		\node [style=empty node] (8) at (0, 1.5) {};
		\node [style=empty node] (9) at (-2.5, 0.25) {};
		\node [style=empty node] (10) at (0, -1) {};
		\node [style=empty node] (11) at (2.5, 0.25) {};
		\node [style=none] (17) at (0, 0.5) {$w_1$};
		\node [style=none] (18) at (-1, 0.75) {$w_0$};
		\node [style=green vertex] (19) at (-1, 0) {};
		\node [style=none] (20) at (1, 0.75) {$w_2$};
	\end{pgfonlayer}
	\begin{pgfonlayer}{edgelayer}
		\draw [style=dashed] (9.center)
			 to [in=-180, out=-90, looseness=1.25] (10.center)
			 to [in=-90, out=0, looseness=1.25] (11.center)
			 to [in=0, out=90, looseness=1.25] (8.center)
			 to [in=90, out=-180, looseness=1.25] cycle;
		\draw (2) to (19);
	\end{pgfonlayer}
\end{tikzpicture}

    &

    \begin{tikzpicture}[auto]
	\begin{pgfonlayer}{nodelayer}
		\node [style=cyan vertex] (1) at (2, 0) {};
		\node [style=green vertex] (2) at (0, 0) {};
		\node [style=empty node] (8) at (0, 1.5) {};
		\node [style=empty node] (9) at (-3, 0) {};
		\node [style=empty node] (10) at (0, -1.5) {};
		\node [style=empty node] (11) at (3, 0) {};
		\node [style=none] (17) at (1, 0.5) {$w_1$};
		\node [style=none] (18) at (-1, 0.5) {$w_0$};
		\node [style=cyan vertex] (19) at (-2, 0) {};
	\end{pgfonlayer}
	\begin{pgfonlayer}{edgelayer}
		\draw (2) to (1);
		\draw [style=dashed] (9.center)
			 to [in=-180, out=-90, looseness=1.25] (10.center)
			 to [in=-90, out=0, looseness=1.25] (11.center)
			 to [in=0, out=90, looseness=1.25] (8.center)
			 to [in=90, out=-180, looseness=1.25] cycle;
		\draw (2) to (19);
	\end{pgfonlayer}
\end{tikzpicture}
        
    \\

    \begin{tikzpicture}[auto, scale=1]
	\begin{pgfonlayer}{nodelayer}
		\node [style=green vertex] (3) at (2, 0) {};
		\node [style=magenta vertex] (0) at (0, 1.25) {};
		\node [style=cyan vertex] (1) at (0, -1.25) {};
		\node [style=green vertex] (2) at (-2, 0) {};
		\node [style=empty node] (8) at (0, 2) {};
		\node [style=empty node] (9) at (-3, 0) {};
		\node [style=empty node] (10) at (0, -2) {};
		\node [style=empty node] (11) at (3, 0) {};
		\node [style=none] (17) at (-0.75, 0) {$w_0$};
		\node [style=none] (18) at (0.75, 0) {$w_1$};
	\end{pgfonlayer}
	\begin{pgfonlayer}{edgelayer}
		\draw [style=filled path] (0.center)
			 to (3.center)
			 to (1.center)
			 to cycle;
		\draw [style=filled path] (0.center)
			 to (2.center)
			 to (1.center)
			 to cycle;
		\draw [style=dashed] (9.center)
			 to [in=-180, out=-90, looseness=1.25] (10.center)
			 to [in=-90, out=0, looseness=1.25] (11.center)
			 to [in=0, out=90, looseness=1.25] (8.center)
			 to [in=90, out=-180, looseness=1.25] cycle;
	\end{pgfonlayer}
\end{tikzpicture}

    &

    \begin{tikzpicture}[auto, scale=1]
	\begin{pgfonlayer}{nodelayer}
		\node [style=cyan vertex] (1) at (1, 0) {};
		\node [style=green vertex] (2) at (-1, 0) {};
		\node [style=empty node] (8) at (0, 1.5) {};
		\node [style=empty node] (9) at (-2, 0) {};
		\node [style=empty node] (10) at (0, -1.5) {};
		\node [style=empty node] (11) at (2, 0) {};
		\node [style=none] (17) at (0, -0.75) {$w_1$};
		\node [style=none] (18) at (0, 0.75) {$w_0$};
	\end{pgfonlayer}
	\begin{pgfonlayer}{edgelayer}
		\draw [bend right] (2) to (1);
		\draw [style=dashed] (9.center)
			 to [in=-180, out=-90, looseness=1.25] (10.center)
			 to [in=-90, out=0, looseness=1.25] (11.center)
			 to [in=0, out=90, looseness=1.25] (8.center)
			 to [in=90, out=-180, looseness=1.25] cycle;
		\draw [bend left] (2) to (1);
	\end{pgfonlayer}
\end{tikzpicture}
    \end{tabular}
    \end{center}

As our terminology suggests, these properties of epistemic coverings are the geometric counterpart of the ones of epistemic frames that we defined previously.

\begin{lemma}
\label{lem:properties-agree}
The properties of \cref{def:frame-properties} agree with the ones of \cref{def:covering-properties} up to the equivalence in \cref{prop:main}.
Namely, if $f$ is a covering of a certain type, then $\kappa(f)$ is of the same type, and conversely for $\sigma$. 
\end{lemma}
\begin{IEEEproof}
    We only show two cases, as the proofs are very similar and just a matter of checking that we correctly translated the notions through the equivalence.
    
    Consider a proper covering $f\!:\!E\to B$.
    In the frame $\kappa(f)$, two worlds $w_1, w_2$
    are indistinguishable by group $\live{w_1}=\live{w_2}$
    if and only if the simplices in $E$ that correspond to $w_1$ and $w_2$
    are sent to the same simplex in $B$.
    But, as $f$ is proper, no two simplices have the same image,
    thus the frame is proper too.
    Now, consider a proper frame $\mathcal{M}$.
    By construction of the functor $\sigma$, two simplices
    $s_1, s_2$ in~$E_M$ of the same color
    are sent to the same simplex in $B_{\sigma(M)}$,
    if and only for corresponding worlds $w_1, w_2$ in $\mathcal{M}$
    $w_1\sim_{\live{w_1}}w_2$.
    But $\mathcal{M}$ is proper, so it is never the case,
    thus $\sigma(\mathcal{M})$ is proper too.

    Consider a covering $f\!:\!E\to B$ with standard group knowledge.
    Suppose that for a pair $w_1, w_2$ in the frame $\kappa(f)$, $w_1\sim_a w_2$ for all $a$ in some $U$. It means that $\partial_a(s_1) = \partial_a(s_2)$, where $s_1$ corresponds to $w_1$ and $s_2$ to $w_2$. As $B$ is a simplicial complex, it follows that $\partial_U(s_1) = \partial_U(s_2)$, so $\kappa(f)$ has standard group knowledge. 
    Take now a frame $M$ with standard group knowledge. Suppose there are two worlds $w_1, w_2$ in $M$ such that $\live{w_1} = \live{w_2} = U$ and for all $a\in U$, $[w_1]_a  = [w_2]_a$. Since $M$ has standard group knowledge, $[w_1]_U = \bigcap_{a\in U} [w_1]_a$. Thus, $[w_1]_U = [w_2]_U$, which means precisely that if two simplices in $B_{\sigma(M)}$ have the same set of vertices, then they are equal.
\end{IEEEproof}

\subsection{Subclasses of epistemic frames and coverings}
\label{sec:subclasses-literature}

The properties of epistemic frames (or, equivalently, epistemic coverings) defined in the previous section can be combined in various ways in order to define particular sub-classes of interest.
The only restriction is that \emph{minimal} and \emph{maximal} are mutually exclusive properties (except in degenerate cases where the model only has empty worlds).
It is also possible, and perhaps sometimes desirable, to consider models that are neither maximal nor minimal.
Moreover, a frame/covering which is \emph{pure} must also be minimal and have no empty world, so there is no need to specify the latter if we already have the former.

We can recover variants of simplicial models that have been defined in previous papers.
Since they are usually concerned with distributed computing, where agents represent processes that can reason about the system, there is little interest in studying empty worlds or empty groups of agents.
Thus, they all have \emph{trivial empty-group knowledge} and \emph{no empty worlds}.
They also have \emph{standard group knowledge} since they work with simplicial complexes instead of semi-simplicial sets.
Moreover, all previous instances of simplicial models were always \emph{proper}, not out of necessity, but because without the notion of covering, it is not possible to model non-proper behaviors in the geometric approach.
On top of that:
\begin{itemize}
\item The original paper introducing simplicial models~\cite{gandalf-journal} was working with \emph{pure} simplicial models (and thus, minimal).
\item In the sequel~\cite{GoubaultLR21kb4}, the ``pure'' assumption is dropped, but models are still assumed to be minimal since formulas are only evaluated at the facets.
\item On the other hand, \cite{Ditmarsch21Wanted} and \cite{Ditmarsch22Complete} study simplicial models that may not be pure but are maximal: every simplex is a possible world.
\item An example that is not concerned about simplicial models, but which studies non-standard group knowledge, is~\cite{baltagCorrelatedKnowledgeEpistemicLogic2010}. Their models are also pure, proper, and have trivial empty-group knowledge.
\end{itemize}
In light of \cref{lem:properties-agree}, we can easily restrict the equivalence of categories proved in \cref{prop:main} to the various sub-categories of frames and coverings. For instance, we can reformulate the main result of~\cite{gandalf-journal} as follows:

\begin{corollary}[\cite{gandalf-journal}]
The category of pure proper epistemic frames with standard group knowledge is isomorphic to the category of pure proper epistemic coverings with standard group knowledge.
\end{corollary}
\begin{IEEEproof}
By \cref{lem:properties-agree}, the restrictions of $\kappa$ and $\sigma$ to those sub-categories is well-defined.
Since we still have $\kappa\circ \sigma = \id$ and $\sigma\circ \kappa = \id$ as proved in \cref{prop:main}, this is still an isomorphism of categories.
\end{IEEEproof}


\section{Axiomatization of the various sub-classes}
\label{sec:axiomatization}

\subsection{Reasoning about alive and dead agents}

As in~\cite{GoubaultLR21kb4}, we can express within the language~$\mathcal{L}_D$ the fact that agents can be dead or alive.
For any agent~$a \in A$ and group of agents~$U \subseteq A$, we define the following formulas: 
\begin{align*}
& \deadprop{a} \,:=\, K_{a} \false
& &
\aliveprop{a} \,:=\, \widehat{K}_{a} \true
\\
& \deadprop{U} \,:=\, \bigwedge_{a \in U} \deadprop{a}
& &
\aliveprop{U} \,:=\, \widehat{D}_U \true
\end{align*}
It is easy to check that they have the expected semantics: for epistemic frames, we have $M,w \models \aliveprop{U}$ iff $w \sim_U w$; and for epistemic coverings, $X,s \models \aliveprop{U}$ iff $s$ is a $V$-simplex with $U \subseteq V$.

\subsection{Axiomatization of epistemic covering models}
\label{sec:Axioms-ECn}

We rely on the usual axiomatization of normal modal logics, with all propositional tautologies,
closure by modus ponens, and the necessitation rule.
On top of that, we add the following five axioms:

\begin{itemize}
    \item ($\mathbf{K}$) $D_U (\varphi \Rightarrow \psi) \Rightarrow (D_U\varphi \Rightarrow D_U \psi)$
    \item ($\mathbf{4}$) $D_U\varphi \Rightarrow D_UD_U\varphi$
    \item ($\mathbf{B}$) $\varphi \Rightarrow D_U\neg D_U\neg\varphi$
    \item ($\mathbf{Mono}$) for $U\subseteq U'$, $D_U\varphi \Rightarrow D_{U'}\varphi$
    \item ($\mathbf{Union}$) for $U,U'$, $\aliveprop{U}\land \aliveprop{U'} \Rightarrow \aliveprop{U\cup U'}$
\end{itemize}

We abbreviate $\mathbf{KB4_n + Mono + Union}$ as $\ECn$, which stands for the logic of epistemic coverings (as we will see).
Notice that the difference between $\mathbf{KB4_n}$ and 
the more standard multi-agent epistemic logic $\mathbf{S5_n}$
is the absence of axiom $\mathbf{T}$: $D_{U}\varphi\Rightarrow\varphi$.
Here are a few examples of valid formulas in $\ECn$ related to the life
and death of agents.

\begin{itemize}
    \item $\ECn \vdash \deadprop{a} \Rightarrow K_a\phi$: dead agents know everything.
    More generally, for $a \in U$,
    $\ECn \vdash \deadprop{a} \Rightarrow D_U\phi$.
    \item $\ECn \vdash \aliveprop{a} \Rightarrow K_a\,\aliveprop{a}$: Alive agents know that they are alive. The same holds for a group~$U$ of agents.
    \item $\ECn \vdash \aliveprop{U} \Rightarrow (D_U\phi \Rightarrow \phi)$: Axiom \textbf{T} is verified when restricted to groups of agents that are alive.
\end{itemize}

We are going to show soundness and completeness of $\ECn$ 
with respect to generalized epistemic models. In order to
show completeness, we use the standard approach:
by providing a canonical model, see for example~\cite{blackburnModalLogic2001}.

\begin{definition}
    \label{def:Mc}
    The \emph{canonical generalized epistemic model} 
    $\Mc = \la \Wc, \sim, L \ra$ is defined as follows:
    \begin{itemize}
    \item $\Wc = \{ \Gamma \mid \Gamma \textup{ is a maximal consistent set of formulas} \}$.
    \item $\Gamma \sim_U \Delta$ \; iff \; $D_U\,\phi \in \Gamma$ implies $\phi \in \Delta$.
    \item $L(\Gamma) = \Gamma \cap \AP$.
    \end{itemize}
\end{definition}

\begin{lemma}[Truth Lemma]\label{lem:truth}
    For any formula $\varphi$ and any maximal consistent set of formulas
    $\Gamma\in M^c$, we have 
    $\varphi\in \Gamma$ iff $M^c,\Gamma\models \varphi$.
\end{lemma}
\begin{IEEEproof}
    We proceed by induction on $\varphi$. 
    The base case of atomic propositions holds by definition of $M^c$. 
    For the boolean connectives, the proof is trivial.

    Let us do the case of $D_U\varphi$.
    Assume that $D_U\varphi$ is in $\Gamma$ and 
    let $\Delta$ be an element of $M^c$ such that $\Gamma\sim_U \Delta$.
    By definition of $\sim$, we have $\varphi\in \Delta$,
    so by induction hypothesis $M^c,\Delta\models \varphi$.
    As $\Delta$ is arbitrarily chosen, we have $M^c,\Gamma\models D_U\varphi$.
    Conversely, assume that $M^c,\Gamma\models D_U\varphi$
    and suppose by contradiction that $D_U\varphi\not\in\Gamma$.
    Then the set 
    $\Delta^- = \{\neg\varphi\} \cup \{\psi\ |\ D_U\psi\in\Gamma\}$ 
    is consistent.
    Indeed, suppose $\Delta^-$ is inconsistent.
    Then we have a proof of 
    $\vdash \psi_1\land\dots\land\psi_k\Rightarrow \varphi$,
    where $D_U\psi_i\in\Gamma$ for every $i$.
    Then, by axiom $\mathbf{K}$, we can prove
    $\vdash D_U\psi_1\land\dots\land D_U\psi_k\Rightarrow D_U\varphi$.
    As $\Gamma$ is maximal consistent, this implies that $D_U\varphi\in \Gamma$,
    which contradicts the assumption.
    Thus, $\Delta^-$ is consistent, and by Lindenbaum's Lemma,
    we can extend it to a maximal consistent set $\Delta\supseteq \Delta^-$.
    By construction, $\Gamma\sim_U \Delta$,
    and by induction hypothesis, $M^c, \Delta \not\models \varphi$.
    This contradicts the assumption that $M^c,\Gamma\models D_U\varphi$.
    Therefore, $D_U\varphi\in \Gamma$, and this concludes the proof.
\end{IEEEproof}

\begin{lemma}\label{lem:alive-canon}
    In the canonical generalized epistemic model $M^c$,
    for any $\Gamma\in M^c$, $U\subseteq\live{\Gamma}$
    iff $\aliveprop{U}\in \Gamma$.
\end{lemma}
\begin{IEEEproof}
    Suppose $U\subseteq \live{\Gamma}$, $\Gamma\sim_{U}\Gamma$.
    Hence, $\Gamma\models \widehat{D}_U\true$, as there is $\Gamma$
    which is $U$-accessible from $\Gamma$, which means that 
    $\aliveprop{U}$ is in $\Gamma$ by the Truth Lemma.
    Conversely, assume $\aliveprop{U}\in \Gamma$. 
    Then, by Truth Lemma, $\Gamma\models \widehat{D}_U\true$,
    so there is $\Delta$, such that $\Delta\sim_{U}\Gamma$.
    By symmetry and transitivity of~$\sim_U$, we have
    $\Gamma\sim_U \Gamma$, so $U\subseteq \live{\Gamma}$.
\end{IEEEproof}

\begin{theorem}\label{thm:complete-simple}
    The logic $\ECn$ is sound and complete with respect to the family of generalized epistemic models.
\end{theorem}
\begin{IEEEproof}
Soundness is straightforward to check: 
the axioms $\mathbf{K}$, $\mathbf{B}$ and $\mathbf{4}$ correspond exactly to the fact that 
the indistinguishability relations are PERs.
Axiom $\mathbf{Mono}$ corresponds to condition~(a) and 
axiom $\mathbf{Union}$ corresponds to condition~(b) in the definition of generalized epistemic frames.

For completeness, consider the canonical model $M^c = \la W^c, \sim, L \ra$ from \Cref{def:Mc}.
Axioms $\mathbf{K,B}$ and $\mathbf{4}$ ensure that $\sim_U$ is a PER,
as in the standard treatment of completeness.
Similarly, axiom $\mathbf{Mono}$ ensures that the generated family of PERs
is monotone: 
assume $U\subseteq U'$, $\Gamma\sim_{U'}\Delta$ and $D_U\varphi\in \Gamma$,
then by $\mathbf{Mono}$, $D_{U'}$ is also in $\Gamma$.
As we assumed $\Gamma\sim_{U'}$, $\varphi\in \Delta$, so $\Gamma\sim_{U}\Delta$.
Axiom $\mathbf{Union}$ ensures that the condition $(b)$ is satisfied.
Assume that $\Gamma\sim_U \Gamma$, $\Gamma\sim_{U'}\Gamma$ 
and $D_{U\cup U'}\varphi\in \Gamma$.
First, we have that if $\Gamma \sim_U \Gamma$, then ${\hat{D}_U\true\in\Gamma}$.
Indeed, by contraposition we have that for any $\psi$, 
$\psi\not\in\Gamma$ entails $D_U\psi\not\in \Gamma$.
In particular, as $\Gamma$ is consistent, $\false\not\in\Gamma$,
so $D_U\false\not\in\Gamma$.
By standard reasoning, $\neg D_U\false\in\Gamma$, as we intended.
Hence, $\aliveprop{U}\land \aliveprop{U'}\in\Gamma$.
By $\mathbf{Union}$ and modus ponens, $\aliveprop{U\cup U'}\in \Gamma$.
As the formula $\aliveprop{V}\Rightarrow (D_V \varphi \Rightarrow \varphi)$ is deducible
when $V=U\cup U'$, we have $D_{U\cup U'}\varphi \Rightarrow \varphi \in \Gamma$.
As we assumed $D_{U\cup U'}\varphi \in \Gamma$, 
by modus ponens $\varphi\in\Gamma$, which shows that the canonical model 
is indeed a generalized epistemic frame.

Applying the Lindenbaum Lemma and the Truth Lemma, 
any consistent formula $\varphi$ holds in some state of $W^c$,
and thus~$\varphi$ is satisfiable. 
\end{IEEEproof}

The following corollary follows from \Cref{lem:equiv-sat}.

\begin{corollary} \label{cor:axiomatization-coverings}
    The logic $\ECn$ is sound and complete with respect to epistemic covering models.
\end{corollary}

\subsection{Axiomatization of other sub-classes}
\label{sec:subclass-axioms}

We now show how to axiomatize the various sub-classes of epistemic models, as discussed in \cref{sec:subclasses-literature}.
As in the previous section, we first establish soundness and completeness for epistemic frames, and then we get the same result for epistemic coverings thanks to~\Cref{lem:equiv-sat,lem:properties-agree}.
The proof is modular: each property of the frames (\cref{def:frame-properties}) corresponds to an extra axiom.
We denote by $U^c$ the complement of the set of agents~$U$, that is $A\setminus U$.

\begin{itemize}
    \item \makebox[1.1cm]{($\mathbf{NE}$) \hfill} $\bigvee_{a\in A} \aliveprop{a}$;
    \item \makebox[1.1cm]{($\mathbf{P}$)\hfill} $\aliveprop{U} \land \deadprop{U^c} \land \varphi \Rightarrow D_U(\deadprop{U^c}\Rightarrow \varphi)$;
    \item \makebox[1.1cm]{($\mathbf{Max}$)\hfill} for $U\not=\emptyset$, $\aliveprop{U} \Rightarrow \neg D_U \neg \deadprop{U^c}$;
    \item \makebox[1.1cm]{($\mathbf{Min}$)\hfill} $\aliveprop{U}\land\deadprop{U^c}\Rightarrow D_U\deadprop{U^c}$;
    \item \makebox[1.1cm]{($\mathbf{Pure}$)\hfill} $\aliveprop{A}$.
\end{itemize}

However, there are no axioms related to trivial empty-group knowledge and standard group knowledge.
This is because those two properties cannot be expressed in the language~$\mathcal{L}_D$.
More formally, we will see in \Cref{thm:complete-max} and \Cref{thm:complete-standard} that $\ECn$ is complete with respect to both classes of structures.
As a consequence, they can be assumed ``for free'', without additional axiom.
Indeed, to show that the logic $\ECn$ is sound and complete with respect to frames with trivial empty-group knowledge, we can use the strategy from \cite{baltagCorrelatedKnowledgeEpistemicLogic2010}.
The proof that standard group knowledge requires no extra axiom is the subject of \Cref{sec:standardgk}.

\begin{remark}
There are several interesting relationships between those axioms.
Axiom $\mathbf{Pure}$, which says that all agents are alive in all worlds, has many consequences. It entails the axioms $\mathbf{NE}$, $\mathbf{Min}$, and $\mathbf{Union}$. Axiom $\mathbf{P}$ is greatly simplified and becomes $\phi \Rightarrow D_A \phi$, where~$A$ is the set of all agents.
Furthermore, $\ECn + \mathbf{Pure}$ together entail Axiom~$\mathbf{T}$, so that the logic $\mathbf{KB4_n}$ becomes~$\mathbf{S5_n}$ when $\mathbf{Pure}$ is assumed.
Another possible interaction is $\mathbf{P} + \mathbf{Min}$, which can be reformulated together as $\aliveprop{U} \land \deadprop{U^c} \land \varphi \Rightarrow D_U\varphi$.
This axiom appears in~\cite{GoubaultLR21kb4}, in the particular case of $U = \{a\}$.
\end{remark}

Now we prove soundness and completeness.

\begin{theorem}\label{thm:complete-max}
    The logic $\ECn \mathbf{+ NE + P + Max}$
    is sound and complete with respect to 
    proper maximal
    epistemic models with 
    trivial empty-group knowledge
    and no empty worlds.
\end{theorem}
\begin{IEEEproof}
For brevity, call an epistemic model with the properties from the theorem
statement a \emph{good maximal} epistemic model.
First, we show soundness.
Suppose we are given a good maximal model $\mathcal{M}$. Axioms of $\ECn$ hold in $\mathcal{M}$ as it is an epistemic model in particular. Axiom $\mathbf{P}$ holds in $\mathcal{M}$ because it is proper: 
assume that $\mathcal{M}, w\models \aliveprop{U} \land \deadprop{U^c}\land \varphi$, that is, $\mathcal{M}, w\models \varphi$ and $\live{w} = U$.
Let $w'\in M$ such that $w'\sim_U w$. Assume then that $\mathcal{M}, w'\models \deadprop{U^c}$, thus $\live{w'} = U$, and by properness of $\mathcal{M}$, $w = w'$ and $\mathcal{M},w' \models \varphi$. For axiom $\mathbf{NE}$, as every world is non-empty, for some $a\in \live{w}$, $K_a\varphi\Rightarrow\varphi$ holds. For axiom $\mathbf{Max}$, suppose for some world $w\in M$, $\mathcal{M}, w\models \aliveprop{U}$, that is $U\subseteq \live{w}$. As $\mathcal{M}$ is maximal, there is always a world $w'$, such that $w'\sim_U w$ and $\live{w'} = U$. So $\neg D_U \neg \deadprop{U^c}$ holds in $w$ since $\deadprop{U^c}$ holds in $w'$.

Now we show completeness. 
We define the canonical model~$M^c$ as in \Cref{def:Mc}, except that consistency now refers to the new logic.
The proof of the Truth Lemma (\Cref{lem:truth}) can easily be adapted.
Let $\varphi_0$ be a consistent formula.
We shall show that $\varphi_0$ is satisfiable in some 
good epistemic model.
Let $M^c$ be the canonical model for the logic,
as in \Cref{thm:complete-simple}.
We need to show that the canonical model is a good model.
Recall that in the canonical model $\Gamma \sim_U \Delta$ iff
for any $\varphi$, $D_U\varphi \in \Gamma \Rightarrow {\varphi\in \Delta}$. The model $M^c$ is:

\paragraph*{Proper}
Suppose $\overline{\Gamma}=\overline{\Delta}=U$ and $\Gamma\sim_U\Delta$.
We need to show that $\Gamma = \Delta$, 
that is for every formula $\varphi$, $\varphi\in \Gamma$ iff $\varphi\in\Delta$.
Suppose $\varphi\in \Gamma$. 
Then, by axiom $\mathbf{P}$, $D_U(\deadprop{U^c}\Rightarrow\varphi)$ is in $\Gamma$.
By the definition of $\sim_U$, it means that $\deadprop{U^c}\Rightarrow\varphi \in \Delta$. By modus ponens, $\varphi\in \Delta$.
Similarly, we can show that $\varphi\in \Delta \Rightarrow \varphi\in \Gamma$.
Hence, $\Gamma = \Delta$ and the canonical model is proper.

\paragraph*{Maximal} 
Let $U\subsetneq \overline{\Gamma}$. 
We want to exhibit a sub-world $\Delta$ of $\Gamma$, 
such that $\overline{\Delta} = U$.
By \Cref{lem:alive-canon}, $\aliveprop{U}\in \Gamma$, so by axiom $\mathbf{Max}$, maximality and consistency,
$\neg D_U \neg \deadprop{U^c}$ is in $\Gamma$.
Then the set 
$\Delta^- = \{ \deadprop{U^c}\} \cup \{\psi\ |\ D_U\psi\in \Gamma\}$
is consistent, by the same reasoning as in the proof of the Truth Lemma.
By Lindenbaum's Lemma, $\Delta^-$ can be extended to 
a maximal consistent set $\Delta$.
Moreover, $\Delta \sim_U \Gamma$ by construction,
so $U\subseteq \live{\Delta}$.
Also, as $\deadprop{U^c}\in \Delta$, $\live{\Delta}\subseteq U$.
Hence, $\live{\Delta} = U$ and $\Delta$ is a sub-world of $\Gamma$.

\paragraph*{No empty worlds} 
By axiom $\mathbf{NE}$, maximality and consistency of any $\Gamma$,
there is an agent $a\in A$, such that $\aliveprop{a}\in \Gamma$.
It entails that $a\in \live{\Gamma}$.

\paragraph*{Trivial empty-group knowledge}
The canonical model $M^c$, however, does not have trivial
empty-group knowledge.
Nevertheless, for every $\varphi_0$, we can extract a sub-model,
$M^c_0$, which consists of the set 
$\{\Gamma\in M^c\ |\ \Gamma\sim_{\varnothing} \Gamma_0\},$
where $\Gamma_0$ is some maximal consistent theory that contains $\varphi_0$.
By monotonicity of $\sim$, this restriction preserves all worlds 
$U$-accessible from $\Gamma_0$ and properties of $\sim$.
Moreover, $M^c_0$ has trivial empty-group knowledge.
Thus, $\varphi_0$ holds in at $\Gamma_0$ in $M^c_0$,
which is proper, maximal, has no empty worlds and trivial empty-group knowledge.
\end{IEEEproof}

Inspecting the proof, one can see that the axioms and properties of
the model are pairwise independent.
Hence, any combination of those axioms yields a sound and complete axiom system for the corresponding class of models.
Another example, with minimal models:

\begin{theorem}
\label{thm:complete-min}
    The logic $\ECn \mathbf{+ NE + P} \mathbf{+ Min }$ 
    is sound and complete with respect to 
    proper minimal
    epistemic models with 
    trivial empty-group knowledge
    and no empty worlds.
\end{theorem}
\begin{IEEEproof}
    Call an epistemic that satisfies properties from the statement of the theorem a good minimal model. The proof of all clauses, except the correspondence between axiom $\mathbf{Min}$ and minimality, is the same as in \Cref{thm:complete-max}. So for soundness, let us show the validity of axiom $\mathbf{Min}$ in any good minimal model.
    Consider a good minimal model $\mathcal{M}$ and a world $w\in \mathcal{M}$. Assume that $\aliveprop{U}\land \deadprop{U^c}$ holds in $w$. It means that $\live{w}$ is exactly $U$. Then by minimality of $\mathcal{M}$, as $w$ is not a subworld of any other world, in any $w'$, such that $w\sim_U w'$, we have $\live{w'}=U$. Hence, $\mathcal{M},w'\models\deadprop{U^c}$ and $\mathcal{M},w\models D_U\deadprop{U^c}$, so axiom $\mathbf{Min}$ is valid.

    For completeness, it remains only to verify that the canonical model $M^c$ for this logic is minimal. Consider $\Gamma, \Delta\in M^c$, such that $\live{\Gamma}\subsetneq\live{\Delta}$ and $\live{\Gamma} = U$. Suppose that $\Gamma\sim_U \Delta$. By \Cref{lem:alive-canon}, $\aliveprop{U}\land\deadprop{U^c}$ belongs to $\Gamma$. By modus ponens and axiom $\mathbf{Min}$, $D_U\deadprop{U^c}$ is in $\Gamma$. By definition of $\sim_U$, $\deadprop{U^c}\in\Delta$. But $\live{\Delta}$ is strictly bigger than $U$, thus there $a\in U^c$ such that $a\in \live{\Delta}$, so $\aliveprop{a}\in \Delta$ which contradicts with consistency of $\Delta$. Thus, $\Gamma\not\sim_U \Delta$ as required.
\end{IEEEproof}

For the case of pure models:
\begin{theorem}
\label{thm:complete-pure}
The logic $\ECn \mathbf{+ NE + P} \mathbf{+ Pure }$ 
is sound and complete with respect to 
proper pure
epistemic models with 
trivial empty-group knowledge
and no empty worlds.
\end{theorem}
\begin{IEEEproof}
    Extending the previous theorem, for soundness we just need to check validity of axiom $\mathbf{Pure}$ in a pure model. Take an epistemic model $\mathcal{M}$ that has properties from the statement of the theorem. The formula $\hat{D}_U\true$ holds in any world $w$ because $w\sim_a w$ as $\mathcal{M}$ is pure.

    For completeness, we need to verify that the canonical model $M^c$ is pure.
    As $\aliveprop{A}$ is in any $\Gamma$ in $M^c$, by \Cref{lem:alive-canon} we have $\live{\Gamma} = A$, so $M^c$ is pure. 
\end{IEEEproof}

\begin{figure}
\begin{center}
\adjustbox{scale=0.6}{%
\begin{tikzcd}[column sep=normal, row sep=0.1em]
        &\parbox{1cm}{\centering$\mathrm{+Max}$ $\mathrm{+NE}$} && \parbox{1cm}{\centering $\mathrm{+Max}$ $\mathrm{+NE}$ $\mathrm{+P}$} \\
        && \parbox{1cm}{\centering$\mathrm{+Max}$ $\mathrm{+P}$} \\
        \mathrm{+Max} & \mathrm{+NE} && \parbox{1cm}{\centering$\mathrm{+NE}$ $\mathrm{+P}$} \\
        && {+\mathrm{P}} \\
        \parbox{1cm}{\centering$\mathrm{KB4}$ $\mathrm{+Mono}$ $\mathrm{+Union}$} & \parbox{1cm}{\centering$\mathrm{+Min}$ $\mathrm{+NE}$} && \parbox{1cm}{\centering$\mathrm{+Min}$ $\mathrm{+NE}$ $\mathrm{+P}$} && \parbox{1cm}{\centering\textcolor{black}{$\mathrm{+Pure}$ $\mathrm{+NE}$ $\mathrm{+P}$}} \\
        && \parbox{1cm}{\centering+$\mathrm{Min}$ $\mathrm{+P}$} \\
        {\mathrm{+Min}}
        \arrow[from=5-1, to=3-1]
        \arrow[from=5-1, to=3-2]
        \arrow[from=5-1, to=4-3]
        \arrow[from=5-1, to=7-1]
        \arrow[from=3-2, to=3-4]
        \arrow[from=4-3, to=3-4]
        \arrow[from=3-1, to=1-2]
        \arrow[from=3-2, to=1-2]
        \arrow[from=4-3, to=2-3]
        \arrow[from=3-1, to=2-3]
        \arrow[from=1-2, to=1-4]
        \arrow[from=2-3, to=1-4]
        \arrow[from=3-4, to=1-4]
        \arrow[from=4-3, to=6-3]
        \arrow[from=7-1, to=6-3]
        \arrow[from=3-2, to=5-2]
        \arrow[from=7-1, to=5-2]
        \arrow[from=5-2, to=5-4]
        \arrow[from=6-3, to=5-4]
        \arrow[from=3-4, to=5-4]
        \arrow[from=5-4, to=5-6]
\end{tikzcd}
}
\end{center}
\caption{Variants of epistemic logic}
\label{fig:modal-cube}
\end{figure}
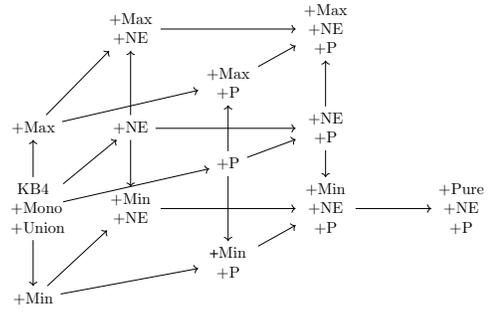

The relationship between all the axiom systems that we consider is summarized in \Cref{fig:modal-cube}.

\subsection{Standard group knowledge}\label{sec:standardgk}

The last property that we did not axiomatize is the standard group knowledge, or, geometrically, the distinction between simplicial sets and simplicial complexes.
In this section, we show that restricting to simplicial complexes does not require any additional axiom.
For that purpose, we use a construction called \emph{unraveling}, which turns a generalized epistemic frame~$M$ into a new frame~$U(M)$ that is bisimilar to~$M$ and has standard group knowledge.

Let $M$ be a generalized epistemic model.
A \emph{history} in $M$ is a finite sequence of the form
$h = (w_0, U_1, w_1, \dots, U_k, w_k)$ for some $k\geq 0$,
such that $w_{i-1}\sim_{U_i} w_i$ for all $1\leq i \leq k$
and $U_i$ is a maximal such subgroup of agents.
Notice that since epistemic models do not necessarily 
have standard group knowledge, every pair of worlds $w, w'$ 
can have several groups~$U$ maximal with respect to inclusion such that 
$w\sim_U w'$. 
For example, if $w\sim_U w'$, $w\sim_{U'} w'$ and both $U,U'$ are maximal,
then one has, in particular, two different histories 
$(w, U, w')$ and $(w, U', w')$. 
We write $\last(h) = w_k$ for the last element of a history,
and we write $h\to_U h'$ if $h' = (h, U, w_{k+1})$ with $U\subseteq U'$.

\begin{definition}
    The \emph{unraveling} of $M$ is a generalized epistemic model 
    $U(M) = (H, \sim^u, L^u)$ defined as follows:
    \begin{itemize}
        \item $\Wu$ is the set of histories of $M$,
        \item $\sim^u_U$ is the transitive and symmetric closure of $\rightarrow_U$, i.e., $\sim^u_U\, = \left( \rightarrow_U \cup \leftarrow_U \right)^*$,
        \item $L^u(h) = L(\last(h))$.
    \end{itemize}
\end{definition}

The unraveling construction is similar to the tree unraveling from \cite{sahlqvistCompletenessCorrespondenceFirst1975}, adjusted to the case of multiple relations related by monotonicity.

\begin{lemma}\label{lem:unravgem}
    Let $M$ be a generalized epistemic model. 
    Then its unraveling $U(M)$ is a generalized epistemic model.
\end{lemma}
\begin{IEEEproof}
    It is easy to see that every $\sim^u_U$ is a PER, as it is 
    transitive and symmetric closure.
    If $U\subseteq U'$, $\rightarrow_{U'}\subseteq \rightarrow_{U}$ 
    by definition, so $\sim^u_{U'}\subseteq \sim^u_{U}$,
    that is $\sim^u$ satisfies the compatibility condition.
    Now, suppose that $h\sim^u_U h$ and $h\sim^u_{U'} h$. 
    By definition of $\sim^u$, it means that there is 
    a sequence of worlds in $M$ such that 
    $\last(h)\sim_U \dots \sim_U \last(h)$,
    so, by transitivity of $\sim_U$, $\last(h)\sim_U \last(h)$.
    Similarly, $\last(h)\sim_{U'}\last(h)$.
    As $M$ is a generalized epistemic model, it follows that
    $\last(h)\sim_{U\cup U'}\last(h)$.
    Thus, there is a history $h' = (h, U\cup U', \last(h))$,
    and $h\sim^u_{U\cup U'} h'$.
    By transitivity and symmetry of $\sim^u_{U\cup U'}$,
    we have $h\sim^u_{U\cup U'} h$, thus concluding that 
    $U(M)$ is closed under union of the groups of alive agents.
\end{IEEEproof}

\begin{lemma}\label{lem:unravstandard}
    For every model $M$, its unraveling $U(M)$ has standard group knowledge.
\end{lemma}
\begin{IEEEproof}
    We need to show that for any two $h, h'$ in $U(M)$,
    if $h\sim^u_U h'$ and $h\sim^u_{U'} h'$, 
    then $h\sim^u_{U\cup U'} h'$.
    There are two cases: if $h=h'$ and $h\not=h'$.
    For the first case, the statement holds since $U(M)$ is a generalized epistemic model by \Cref{lem:unravgem}.

    For the second case, notice first that $\to_V$ respects the ordering
    of histories with respect to the prefix relation: 
    if $h\to_V h'$, then $h$ is a prefix of $h'$. 
    The prefix relation forms a tree on the set of histories,
    and it implies that if $h\sim^u_V h'$, then 
    there is a unique non-redundant path
    $h\leftarrow_V \dots \leftarrow_V h'' \rightarrow_V \dots \rightarrow_V h'$ from $h$ to $h'$ that witnesses it, 
    where $h''$ is the common prefix of $h$ and $h'$.
    Moreover, this path is the same for any $V\subseteq A$.
    We can write $h = (h'', V_1, w_1,\dots,V_n, w_n)$ and 
    $h' = (h'', V'_1, w'_1, \dots, V'_m, w_m)$.
    As it is the same path for both $U$ and $U'$,
    $U\cup U'\subseteq V_i$ and $U\cup U'\subseteq V'_j$ for all $i,j$.
    In particular, we have that 
    $\last(h)=w_n\sim_{U\cup U'} \dots \sim_{U\cup U'} \last(h'') \sim_{U\cup U'} \dots \sim_{U\cup U'} w'_m = \last(h')$.
    Thus, $h\sim^u_{U\cup U'} h'$, and $U(M)$ has standard group knowledge.
\end{IEEEproof}

\begin{remark}\label{rem:lastforth}
    In the proof we have also shown that if ${h\sim^u_U h'}$,
    then $\last(h)\sim_U \last(h')$.
\end{remark}

Given two generalized epistemic models $M, N$, 
we say that morphism $p: M\to N$ is a \emph{functional bisimulation}
if the following conditions hold:
\begin{itemize}
    \item (atoms) for any $w\in M$, $L_N(f(w)) = L_M(w)$;
    \item (forth) for all $U\subseteq A$, 
    if $w\sim^M_U w'$, then $f(w)\sim_U f(w')$;
    \item (back) for all $U\subseteq A$, if $f(w)\sim^N_U v'$, 
    then there is $w'\in M$ such that $f(w')=v'$ and $w\sim^M_U w'$.
\end{itemize}

The definition we give is an extension of the standard notion of bisimulation, 
which links structural similarity of models with validity of formulas.
In particular, we have the following proposition by adapting the standard construction (see \cite{blackburnModalLogic2001}).

\begin{proposition}
    If $f: M\to N$ is a functional bisimulation, then for any formula $\varphi$,
    we have that $M, w\models \varphi$ if and only if $N, f(w)\models \varphi$.
\end{proposition}

This is the essential proposition that allows us to show that $\ECn$ is 
sound and complete with respect to frames with standard group knowledge.

\begin{lemma}\label{lem:lastbisim}
    For every generalized epistemic model $M$, its unraveling $U(M)$ is bisimilar to $M$.
\end{lemma}
\begin{IEEEproof}
    We shall show that $\last: U(M)\to M$ is a functional bisimulation.
    First, by the definition of $L^u$, $L^u(h) = L(\last(h))$, 
    so the atomic proposition are preserved.

    Suppose now that $h\sim^u_U h'$. 
    By \Cref{rem:lastforth}, $\last(h)\sim_U h'$, 
    so (forth) condition is satisfied.

    For (back) condition, suppose that $\last(h)\sim_U w'$.
    Then there is a history $h' = (h, U', w')$, such that $U\subseteq U'$.
    Clearly, $h \to_U h'$, so $h\sim^u_U h'$, and the (back) condition holds too, thus concluding that $\last: U(M)\to M$ is a bisimulation.
\end{IEEEproof}

\begin{theorem}
\label{thm:complete-standard}
    The logic $\ECn$ is sound and complete with respect to models with standard group knowledge.
\end{theorem}

\begin{IEEEproof}
    Soundness is straightforward, as models with standard group knowledge are generalized epistemic models in particular.

    Consider the canonical model $M^c$ for $\ECn$ 
    from \cref{thm:complete-simple}.
    It is shown that $M^c$ is a generalized epistemic frame.
    By \Cref{lem:unravgem}, $U(M^c)$ is a generalized epistemic frame,
    and by \Cref{lem:unravstandard} it has standard group knowledge.
    For any formula $\varphi$, there is $\Gamma$ in $M^c$, 
    such that $M^c, \Gamma\models \varphi$.
    Since $\last: U(M^c)\to M^c$ is a bisimulation by \Cref{lem:lastbisim},
    for any $h\in U(M^c)$ such that $\last(h) = \Gamma$, we have $U(M^c), h\models \varphi$. Thus, any formula $\varphi$ is valid in a model with classic group knowledge, which concludes completeness.
\end{IEEEproof}

\subsection{Topological interpretation}

The completeness result with respect to models with standard group knowledge points us towards the study of expressivity of modal logic with respect to topological properties of csets. In particular, a natural question is what kind of similarity between csets is induced by a bisimulation? In this subsection, we give a partial answer to this question. 

Let us restrict our attention to good maximal epistemic models.
By the isomorphism we have shown, it is the same as to consider csets where every simplex is a unique world. Let $\alpha: X \to Y$ be a functional bisimulation.
We can translate the definition of a bisimulation in terms of epistemic coverings: it means that the valuation of a simplex $s\in X$ coincides with the valuation of its image $\alpha(s)\in Y$; if two simplices $s,s'$ in $X$ have a common $U$-subsimplex, then $\alpha(s)$ and $\alpha(s')$ have a common $U$-subsimplex;
if $\alpha(s)$ has a common $U$-subsimplex with $t'$ in $Y$, then there is a simplex $s'$ in $X$ that shares a $U$-subsimplex with $s$, and $\alpha(s') = t'$.
The second condition is always satisfied by any morphism of csets.
The last condition is more interesting, as it can be interpreted as a certain type of \emph{lifting condition} that is ubiquitous in topology.
This point of view can be traced back to \cite{joyalBisimulationOpenMaps1996}.

	
Without giving a proof, we remark that the forgetful functor from csets to simplicial sets (which just forgets colors of vertices) sends functional bisimulations of csets with \emph{unique} lifting (corresponding to having a unique simplex $s'$ with $\alpha(s')=t'$ in the condition above) to \emph{simplicial coverings} (see \cite{gabrielCalculusFractionsHomotopy1967}). An example of such a bisimulation that can be seen as a topological covering is depicted. It represents the classical example of a covering of a circle by an infinite helix.

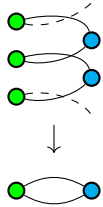
\begin{center}
    \begin{tikzpicture}
	\begin{pgfonlayer}{nodelayer}
		\node [style=green vertex] (0) at (-1, -1) {};
		\node [style=cyan vertex] (1) at (1, -1) {};
		\node [style=green vertex] (2) at (-1, 1.5) {};
		\node [style=green vertex] (3) at (-1, 2.5) {};
		\node [style=green vertex] (4) at (-1, 3.5) {};
		\node [style=cyan vertex] (5) at (1, 2) {};
		\node [style=cyan vertex] (6) at (1, 3) {};
		\node [style=none] (7) at (1, 1) {};
		\node [style=none] (8) at (1, 4) {};
		\node [style=none] (9) at (0, 0.75) {};
		\node [style=none] (10) at (0, 0) {};
	\end{pgfonlayer}
	\begin{pgfonlayer}{edgelayer}
		\draw [bend right] (0) to (1);
		\draw [bend left] (0) to (1);
		\draw [in=-135, out=-15, looseness=0.75] (2) to (5);
		\draw [in=15, out=105, looseness=0.75] (5) to (3);
		\draw [in=-135, out=-15, looseness=0.75] (3) to (6);
		\draw [in=15, out=105, looseness=0.75] (6) to (4);
		\draw [style=dashed path, in=-135, out=-15, looseness=0.75] (4) to (8.center);
		\draw [style=dashed path, in=120, out=15, looseness=0.75] (2) to (7.center);
		\draw [style=arrow path] (9.center) to (10.center);
	\end{pgfonlayer}
\end{tikzpicture}
\end{center}

\section{Conclusions and future work}

We have introduced a very general class of epistemic models based on generalized epistemic frames, and their geometric counterpart based on simplicial sets, epistemic covering models.
These models subsume many variants of simplicial models found in the literature.
We made good use of this generality by establishing a close connection between properties of the models and axioms of the logic.
This yields soundness and completeness for a variety of logics (\cref{fig:modal-cube}).

A notable fact that we prove is that every model based on simplicial sets is equivalent modulo bisimulation to a model based on simplicial complexes.
This unveils interesting geometric considerations: epistemic modal logics do not distinguish simplicial sets from their coverings because of the local nature of the knowledge operators.
Could we define a logic that is able to capture better the global geometry of the model?

In future work, we aim at putting these logics in action for, in particular, distributed computing applications such as in e.g.\ \cite{gandalf-journal,Ditmarsch2020KnowledgeAS,GoubaultLLR19disc,Nishimura22mucalculus}, and bridge the gap between the geometric interpretations \cite{coverageGhrist} of distributed problems such as sensor networks, as touched upon in Example \ref{ex:sensnet}, with their logical interpretations. 


\providecommand{\noopsort}[1]{}

\end{document}